\documentclass[]{aa}

\usepackage{txfonts,epsfig,graphicx,natbib,url,twoopt}
\usepackage{import}
\usepackage{paralist}
\usepackage{booktabs,xspace}
\usepackage{amsmath}
\usepackage{tikz}
\usepackage{amsmath}
\usepackage{amssymb}
\usepackage{graphicx}
\usepackage{float}
\usepackage[export]{adjustbox}

\bibpunct{(}{)}{;}{a}{}{,}

\usepackage[hidelinks=true,breaklinks=true]{hyperref} 
\usepackage{natbib,twoopt}
\bibpunct{(}{)}{;}{a}{}{,} 
\makeatletter
 \newcommandtwoopt{\citeads}[3][][]{%
   \href{http://adsabs.harvard.edu/abs/#3}%
        {\def\hyper@linkstart##1##2{}%
         \let\hyper@linkend\@empty\citealp[#1][#2]{#3}}
	}
 \newcommandtwoopt{\citepads}[3][][]{%
   \href{http://adsabs.harvard.edu/abs/#3}%
        {\def\hyper@linkstart##1##2{}%
         \let\hyper@linkend\@empty\citep[#1][#2]{#3}}
	}
 \newcommandtwoopt{\citetads}[3][][]{%
   \href{http://adsabs.harvard.edu/abs/#3}%
        {\def\hyper@linkstart##1##2{}%
         \let\hyper@linkend\@empty\citet[#1][#2]{#3}}
	}
 \newcommandtwoopt{\citeyearads}[3][][]{%
   \href{http://adsabs.harvard.edu/abs/#3}%
        {\def\hyper@linkstart##1##2{}%
         \let\hyper@linkend\@empty\citeyear[#1][#2]{#3}}
	}
\makeatother

\renewcommand{\vec}{\boldsymbol} 

\begin{document}
	\title{Gaia astrometry for stars with too few\\observations -- a Bayesian approach}
	\author{
		Daniel Michalik\inst{1}
			\and
		Lennart Lindegren\inst{1}
			\and
		David Hobbs\inst{1}
			\and
		Alexey~G. Butkevich\inst{2}
	}

	\institute{
		Lund Observatory, Department of Astronomy and Theoretical Physics, Lund University, Box 43, 22100 Lund, Sweden\\
		\email{[daniel.michalik; lennart; david]@astro.lu.se}
		\and
		Lohrmann Observatory, Technische Universit\"at Dresden,
                01062 Dresden, Germany\\
		\email{alexey.butkevich@tu-dresden.de}
	}

	\date{Accepted for publication in \emph{Astronomy and Astrophysics}, 2015 August 21}

	\abstract
{The astrometric solution for Gaia aims to determine at least five parameters
for each star, representing its position, parallax, and proper motion, together
with appropriate estimates of their uncertainties and correlations. This requires at least five
distinct observations per star. In the early data reductions the number of
observations may be insufficient for a five-parameter solution, and even after the full mission many stars
will remain under-observed, including faint stars at the detection limit and
transient objects. In such cases it is reasonable to determine only the two position
parameters. The formal uncertainties of such a two-parameter solution would
however grossly underestimate the
actual errors in position, due to the neglected parallax and proper motion.}
{We aim to develop a recipe to calculate sensible formal uncertainties that
can be used in all cases of under-observed stars.
}
{ Prior information about the typical ranges of stellar parallaxes and proper
motions is incorporated in the astrometric solution by means of Bayes' rule.
Numerical simulations based on the Gaia Universe Model Snapshot (GUMS) are used to
investigate how the prior influences the actual errors and formal uncertainties when
different amounts of Gaia observations are available. We develop a criterion for the
optimum choice of priors, apply it to a wide range of cases, and derive a
global approximation of the optimum prior as a function of magnitude and
galactic coordinates.
}
{
The feasibility of the Bayesian approach is demonstrated through global
astrometric solutions of simulated Gaia observations. With an appropriate
prior it is possible to derive sensible positions with realistic
error estimates for any number of available observations. Even though this
recipe works also for well-observed stars it should not be used where a good
five-parameter astrometric solution can be obtained without a prior. Parallaxes and
proper motions from a solution using priors are always biased and should
not be used.
}
{}

\maketitle
%

\section{Motivation for this study\label{sec:introduction}}

The ESA science mission Gaia, launched in December 2013, aims to determine
accurate astrometry (positions, parallaxes, and proper motions) and complementary
spectrophotometry for about one billion stars \citep{2001A&A...369..339P,2012Ap&SS.341...31D}.
The astrometric parameters of a given star are calculated from
the transits of the star's image across the CCDs in the focal plane of Gaia.
Each such field-of-view transit is essentially an instantaneous, one-dimensional
measurement of the stellar position in a certain scan direction (the `along-scan coordinate').
The perpendicular (`across-scan') coordinate is also measured, but to a lower
accuracy, and does not contribute significantly to the final astrometric parameters.

The path of a star on the celestial sphere, as seen from Gaia, is in the simplest
case modelled by five astrometric parameters representing its position
($\alpha$, $\delta$), parallax ($\varpi$), and proper motion ($\mu_{\alpha*}$,
$\mu_\delta$) at some chosen reference epoch.
To determine all five parameters one needs at least five observations suitably
distributed in time, and different scan directions are needed to derive the two-dimensional
positions from the one-dimensional scans. Due to the one-year periodicity of parallax,
the observations must span at least a whole year in order to reliably disentangle parallax
from proper motion. The Gaia scanning law ensures that these
conditions are met for stars anywhere in the sky, if the scanning lasts long enough.
The nominal mission length of five years provides an ample number of observation
opportunities, with an average of some 70 field-of-view transits per star. This high
redundancy factor is needed to determine a large number of nuisance parameters
(attitude and instrument calibration) in addition to the
astrometric parameters, for judging the quality of the data, and for detecting cases
(such as binaries) where the simple five-parameter model is not adequate.

However, there are inevitably many situations where a star is insufficiently
observed to solve all of its five astrometric parameters. These situations include:
\begin{itemize}
\item Transient objects, for example extragalactic supernovae, galactic dwarf novae,
and large-amplitude (Mira type) variables: these may be visible for just a few
months, possibly reoccurring at a much later date.
\item Faint stars near the detection limit of Gaia: nominally, all point
sources brighter than 20th magnitude are detected and observed.  However, since
the on-board magnitude estimation has some uncertainty, stars at the detection
limit may not always be observed when they transit the focal plane. Because the
detection probability decreases gradually with magnitude, large numbers of
faint stars will have strongly diluted observation histories.
\item The first release of astrometric results, based mainly on observations
collected during the first year of the mission, where most stars will be
insufficiently observed.
\end{itemize}
If there are not enough observations for a given star, a simple remedy is to solve only
its position ($\alpha$, $\delta$) at the mean epoch of observation. This is always
possible: even in the case of a single field-of-view transit, an approximate position
can be calculated by combining the along-scan and across-scan measurements.

Solving only for the two position parameters $\alpha$ and $\delta$ is equivalent
to assuming that the true parallax and proper motion of the object are equal to zero.\footnote{
We do not consider the possibility of solving three or four astrometric parameters
per star, for example ($\alpha$, $\delta$, $\varpi$) or  ($\alpha$, $\delta$,
$\mu_{\alpha*}$, $\mu_\delta$). This would mean that proper motion is neglected
compared to parallax, or vice versa. This makes little sense because, for most stars,
the observable effect of the neglected parameter will be of a similar size
as that of the retained parameter. This follows from the speed of the
Earth's motion around the Sun, about 30~km~s$^{-1}$, being of a similar magnitude
as the peculiar motions of stars, including that of the Sun itself. Consequently
we only consider solutions with either two or five astrometric parameters per star.}
If this assumption is correct (as may effectively be the case e.g.\ for quasars), the
resulting position estimate will be unbiased with a formal uncertainty
reflecting the actual
errors. However, if the true parallax and proper motion are non-zero, the estimated
position will in general be biased. Its formal uncertainty (which does not depend
on the parallax and proper motion value) will remain small,
since the error calculus only takes into account the small observational noise of
Gaia. As a result, the bias will often be many times larger than the
formal uncertainty.

The solution proposed in this paper is to estimate all five parameters, while
incorporating the prior information that the parallax and proper motion are typically
small but non-zero quantities. Formally, this can be achieved by means of Bayes' rule.
This paper tries to answer the question how to optimally choose the prior when
there are not enough Gaia observations for a regular five-parameter astrometric
solution. We use numerical experiments, based on simulated observations of
stars in a galactic model, to investigate the influence of the prior under
different scenarios. We show that with a suitable choice of prior the solution
provides sensible results in terms of both the estimated position and its
calculated uncertainty.

The first release of astrometric results from the Gaia mission is expected\footnote{See \url{http://www.cosmos.esa.int/web/gaia/release}.}
 in the summer
of 2016. Due to the limited time interval covered by the early data, this release will,
for the majority of stars, only contain mean positions and single-band
($G$) magnitudes. Exceptions are the Hipparcos stars, for which improved proper
motions and possibly also parallaxes can be derived based on the HTPM project
\citep{LL:FM-040,2014A&A...571A..85M}.
A similar joint reduction is possible for the Tycho-2 stars (the TGAS project; \citealt{2015A&A...574A.115M}).

The method developed in this paper could be applied to the estimation of the positional
uncertainties in the first release, but more generally to any situation where the number
and distribution of observations is insufficient for a full five-parameters solution.
It should be emphasised that the use of prior information in the astrometric solution
always leads to biased estimates of the parameters. The proposed recipe
should therefore only be used when actually needed, e.g.\ in the previously mentioned
cases, and then only in order to obtain positions with realistic estimates of their
uncertainties. These positions are valuable, e.g.\ for identification purposes
and as a reference for ground-based observations. The resulting parallaxes and proper
motions should however not be used.

\section{Theory\label{sec:theory}}

In this section we first formulate the estimation of the astrometric parameters
as a classical least-squares problem, which provides a connection to the
description of the overall Gaia astrometric solution \citep{2012A&A...538A..78L}.
We then show how a Gaussian prior can be introduced using Bayes' rule.
Finally we discuss the relevance and interpretation of the Gaussian prior
and posterior probability densities in this context.

We use the term uncertainty for any quantitative measure of the expected degree
of deviation of an estimated quantity from its true value, and reserve the term
(actual) error for the signed, and in general unknown, deviation itself. In the Gaussian
context the natural measure of uncertainty is the standard deviation, but as
we are here dealing with strongly non-Gaussian distributions (e.g.\ of the true parallax values) we instead use measures based
on the size of a confidence region.

\subsection{Least-squares estimation of the astrometric parameters\label{sec:leastsquare}}
The Gaia astrometric solution is calculated by a series of updating processes
as described in Sect.~5 of \citet{2012A&A...538A..78L}. In the `astrometric updating'
the satellite attitude and geometric calibration are assumed to be known, in
which case the linearised least-squares problem for an individual star can be
written in matrix form as
\begin{equation}\label{eq:Axh}
\vec{A} \vec{x} \simeq \vec{h}\, ,
\end{equation}
where $\vec{x}$ is a column matrix containing differential corrections
to the five astrometric parameters, $\vec{h}$ is a column matrix containing
the $n$ pre-adjustment observation residuals of the star, normalized by their formal uncertainties,
and $\vec{A}$ is the $n\times 5$ design matrix, i.e., the partial derivative matrix
row-wise normalized by the observational uncertainties. (The $\simeq$ is used
in Eq.~\ref{eq:Axh} because the system of equations is in general overdetermined
and cannot be exactly satisfied.)

The astrometric parameters $\alpha$, $\delta$, $\varpi$,
$\mu_{\alpha *}\equiv\mu_\alpha \cos\delta$ and $\mu_\delta$ refer to
some chosen reference epoch $t_\text{ep}$, which in this paper is always
taken to be the mean epoch of observation.  In particular, $(\alpha,\delta)$ is
the barycentric direction to the star at time $t_\text{ep}$. The differential
corrections in $\vec{x}$ should be interpreted as
$\Delta\alpha*\equiv\Delta\alpha\cos\delta$, $\Delta\delta$, $\Delta\varpi$,
$\Delta\mu_{\alpha*}$, and $\Delta\mu_\delta$, or more rigorously using
the `scaled modelling of kinematics' formalism in Appendix~A of
\citet{2014A&A...571A..85M}.%
\footnote{The rigorous treatment includes the radial proper motion $\mu_r$
as the sixth astrometric parameter. Even for nearby high-velocity stars the
perspective effect in position, which is proportional to $\mu_r$, is negligible over five
years. In the present problem $\mu_r$ can therefore be ignored.}

The least-squares estimate of $\vec{x}$ minimizes the $\chi^2$ goodness-of-fit, i.e., the
squared norm of the post-fit residuals $\vec{h} - \vec{A}\vec{x}$,
\begin{align}\label{eq:q0}
Q_0(\vec{x}) &= \lVert \vec{h} - \vec{A}\vec{x} \rVert^2 = (\vec{h} - \vec{A}\vec{x})'(\vec{h} - \vec{A}\vec{x})\nonumber\\
	&= \vec{h}'\vec{h} - 2\vec{x}'\vec{b}_0 + \vec{x}'\vec{N}_0\vec{x}\, ,
\end{align}
where $\vec{b}_0 = \vec{A}'\vec{h}$ and $\vec{N}_0 = \vec{A}'\vec{A}$.
Putting $\partial Q_0 / \partial \vec{x} = \vec{0}$ gives a linear system of equations,
\begin{equation}
\vec{N}_0\vec{x}_0 = \vec{b}_0\, ,
\end{equation}
known as the normal equations, from which the least-squares estimate $\vec{x}_0$
can be calculated.
For arbitrary $\vec{x}$ the goodness-of-fit can be written as
\begin{equation}\label{eq:q0a}
Q_0(\vec{x}) = Q_0(\vec{x}_0) + (\vec{x} - \vec{x}_0)' \vec{N}_0 (\vec{x} - \vec{x}_0) \, .
\end{equation}

For badly observed stars the normal matrix $\vec{N}_0$ will be either ill-conditioned
or singular. If it is ill-conditioned (e.g., due to a small number of nearly
collinear observations), then a solution can formally be obtained. It will
however have large formal uncertainties and be vulnerable to outliers, which
cannot be reliably detected.
The situation is different if $\vec{N}_0$ is strictly singular, e.g., if
there are fewer observations than the number of unknowns. From a mathematical point of
view, the singular problem possesses an infinite number of solutions, while
algorithmically it may not be possible to determine any of them, depending on
implementation choices. A remedy to both singular and ill-conditioned
situations is to incorporate prior information (Sect~\ref{sec:incorporating}),
which always results in a unique and well-defined, albeit biased, solution.

\subsection{The likelihood function}

For a clean data set, with outliers filtered out or downweighted, it is reasonable to model
the observational errors as independent normal random variables. For a properly
calibrated instrument, the errors have mean (expected) values equal to zero and known
standard deviations equal to the formal uncertainties of the observations.
$\vec{h}$ is then an $n$-dimensional Gaussian, with mean value $\vec{A}\vec{x}_\text{true}$
and unit covariance; its probability density function (PDF) is
\begin{equation}
f(\vec{h}|\vec{x})=(2\pi)^{-n/2} \exp\left[-{\textstyle\frac{1}{2}} \lVert\vec{h}-\vec{A}\vec{x}\rVert^2\right]
\propto \exp\left[-{\textstyle\frac{1}{2}} Q_0(\vec{x})\right]\, ,\label{eq:likelihood}
\end{equation}
evaluated for $\vec{x}=\vec{x}_\text{true}$. Naturally, this PDF cannot be computed as
$\vec{x}_\text{true}$ is unknown. Regarded as a
function of $\vec{x}$, for the given $\vec{h}$, it is known as the likelihood of the data,
designated $L(\vec{x}|\vec{h})$.
Maximizing this function with respect to $\vec{x}$ is clearly equivalent to minimizing
$Q_0(\vec{x})$, showing that $\vec{x}_0$ is the maximum likelihood estimate of
the astrometric parameters.

\subsection{Incorporating a prior\label{sec:incorporating}}

Bayes' rule (e.g.\ \citealt{Sivia2006}, Sect.~3.5) expresses the posterior PDF
of $\vec{x}$ as
\begin{equation}
f(\vec{x}|\vec{h}) \propto L(\vec{x}|\vec{h}) \times p(\vec{x})\, ,\label{eq:bayes}
\end{equation}
where $p(\vec{x})$ is the prior PDF and $L(\vec{x}|\vec{h})\equiv f(\vec{h}|\vec{x})$ is the likelihood of the
data.
The constant of proportionality is left out as it is independent of $\vec{x}$, but can be
determined from the normalization constraint $\int f(\vec{x}|\vec{h})\,\text{d}\vec{x}=1$.
For example, a flat (uninformative) prior $p_0(\vec{x})=\text{const}$ yields, by
means of Eqs.~(\ref{eq:q0a})--(\ref{eq:likelihood}), the posterior PDF
\begin{equation}
f_0(\vec{x}|\vec{h}) = (2\pi)^{-5/2}\det(\vec{N}_0)^{1/2}\exp\left[-{\textstyle\frac{1}{2}} Q_0(\vec{x})\right]\, .
\label{eq:posterior0}
\end{equation}
This is a 5-dimensional Gaussian with mean value $\vec{x}_0$ and covariance
$\vec{C}_0 = \vec{N}_0^{-1}$, which reflects our knowledge of $\vec{x}$ based
on the data only.

In principle, the prior PDF $p(\vec{x})$ should quantify our prior knowledge of the
astrometric parameters. For example, it could be
strictly zero for $\varpi<0$, while declining as a power law for large values of
$\varpi$, reflecting the prior knowledge that parallaxes are generally positive, small quantities. However, in this paper we only consider Gaussian priors.
This has two important advantages: (a) if both the prior PDF and the likelihood
function are Gaussian, the posterior PDF is also Gaussian, which greatly simplifies
its interpretation; (b) the incorporation of a Gaussian prior in the astrometric solution
is straightforward, as will be shown in the following. The disadvantage is of
course that a Gaussian prior is not very realistic, at least for the parallax; but
with the interpretation proposed in Sect.~\ref{sec:interpretation} it is
adequate for the present purpose.

Assuming a Gaussian prior with mean value
$\vec{x}_p$ and covariance $\vec{C}_p$ we define
\begin{equation}
Q_p(\vec{x}) = (\vec{x} - \vec{x}_p)' \vec{N}_p (\vec{x} - \vec{x}_p)\,,
\end{equation}
where $\vec{N}_p = \vec{C}^{-1}_p$.
The prior probability density function is then
\begin{equation}
p(\vec{x}) \propto \exp\left[-{\textstyle\frac{1}{2}}Q_p(\vec{x})\right]\,.\label{eq:priorpdf}
\end{equation}
Inserting Eqs.~(\ref{eq:likelihood}) and (\ref{eq:priorpdf}) into Eq.~(\ref{eq:bayes}) yields
the posterior PDF
\begin{equation}\label{eq:posterior}
f(\vec{x}|\vec{h}) \propto \exp\left[ - {\textstyle\frac{1}{2}} Q_0(\vec{x}) -{\textstyle\frac{1}{2}} Q_p(\vec{x})\right] \, .
\end{equation}
Being the product of two Gaussian distributions, $f(\vec{x}|\vec{h})$ 
is clearly also Gaussian. The expected value of $\vec{x}$ can therefore be obtained by minimizing
$Q(\vec{x}) = Q_0(\vec{x}) + Q_p(\vec{x})$, i.e., by solving
\begin{align}
\partial Q(\vec{x}) / \partial \vec{x} &= 2 N_0(\vec{x} - \vec{x}_0) + 2 N_p (\vec{x} - \vec{x}_p) = 0\, ,
\end{align}
or
\begin{align}
(\vec{N}_0 + \vec{N}_p) \vec{x} = \vec{b}_0 + \vec{b}_p\,,\label{eq:priorincoporation}
\end{align}
where
\begin{align}
\vec{b}_p = \vec{N}_p \vec{x}_p\,.\label{eq:priorincoporation2}
\end{align}
It is readily shown that the
covariance of the posterior estimate is given by
\begin{align}
\vec{C} = (\vec{N}_0 + \vec{N}_p)^{-1} \, . \label{eq:posteriorC}
\end{align}
Equations~(\ref{eq:priorincoporation})--(\ref{eq:posteriorC}) are the theoretical basis for
the `joint solution' scheme of incorporating {\sc Hipparcos} and Tycho-2 priors in the Gaia data processing, developed
for the HTPM and TGAS projects \citep{2014A&A...571A..85M,
2015A&A...574A.115M}. In the following the prior is not derived from earlier
catalogues but from our expectation of the distributions of parallaxes and
proper motions.

\subsection{Interpretation of the Gaussian probability densities}
\label{sec:interpretation}

In the following it is assumed that the prior distribution of
parallaxes is Gaussian with mean value $\varpi_p=0$ and standard deviation
$\sigma_{\varpi,p}$ equal to the square root of the corresponding (third)
diagonal element of $\vec{C}_p$. (Similar assumptions are made concerning
the prior distributions of the proper motion components.)
Clearly this is not very realistic, as it
implies that, a~priori, there is a 50\% probability that the parallax
is negative. However, the same Gaussian prior also means that there is a
90\% probability that the true parallax is less than $1.28\sigma_{\varpi,p}$,
and a 99\% probability that it is less than $2.33\sigma_{\varpi,p}$.
These latter statements are obviously meaningful, and provide a useful
quantification of the expected \emph{smallness} of the parallax, even
though the distribution of true parallaxes is far from Gaussian.

A similar interpretation can be made of the Gaussian posterior PDF in
Eq.~(\ref{eq:posterior}). Although the actual error distribution of the
Bayesian solution may be strongly non-Gaussian, this PDF can
still be used to construct sensible confidence regions. In this work we
are primarily interested in the positions and ignore the estimated parallaxes
and proper motions. As the position uncertainty may be quite anisotropic,
it should not be given as a single value but as a confidence region,
for example a confidence ellipse, such that the true value is contained
within that region with a certain degree of confidence $P$.

In this work we choose to work with a confidence level of 90\% ($P=0.9$).
This means that the (Gaussian) posterior covariance should be such that a
90\% confidence ellipse constructed from it will, in 90\% of the cases, contain
the true position. The choice of $P=0.9$ is arbitrary, and a different value
(e.g., 0.8, 0.95, or 0.99) would in general require a different covariance
matrix in order to correctly characterise the errors at that $P$-value. Only in
the case of Gaussian posterior errors would a single covariance matrix
correctly describe the error distribution for different values of $P$.

The confidence ellipse can be constructed from the positional covariance
(the $2\times 2$ submatrix of $\vec{C}$) as described in
\citet{Press:2007:NRE:1403886}. In particular, for $P=0.9$ the semi-axes
of the ellipse are $\sqrt{-2 \ln(1-P)}\simeq 2.146$ times the square roots
of the singular values of the positional covariance matrix. Inside the ellipse
we have $Q(\vec{x})-Q_\text{min}<-2\ln(1-P)\simeq 4.605$.

A good astrometric solution should not only be as accurate as possible but
also have formal uncertainties that characterize the actual errors correctly.
Thus our general approach is to optimise the prior PDF for both goals.
The formal uncertainties (positional covariance matrix) of the resulting
posterior estimate should be such that the 90\% confidence ellipse, computed
as described above, contains the true position with 90\% probability.

\begin{figure}[t]
\centering
\includegraphics[width=.35\textwidth,clip,trim=0 0 20 0]{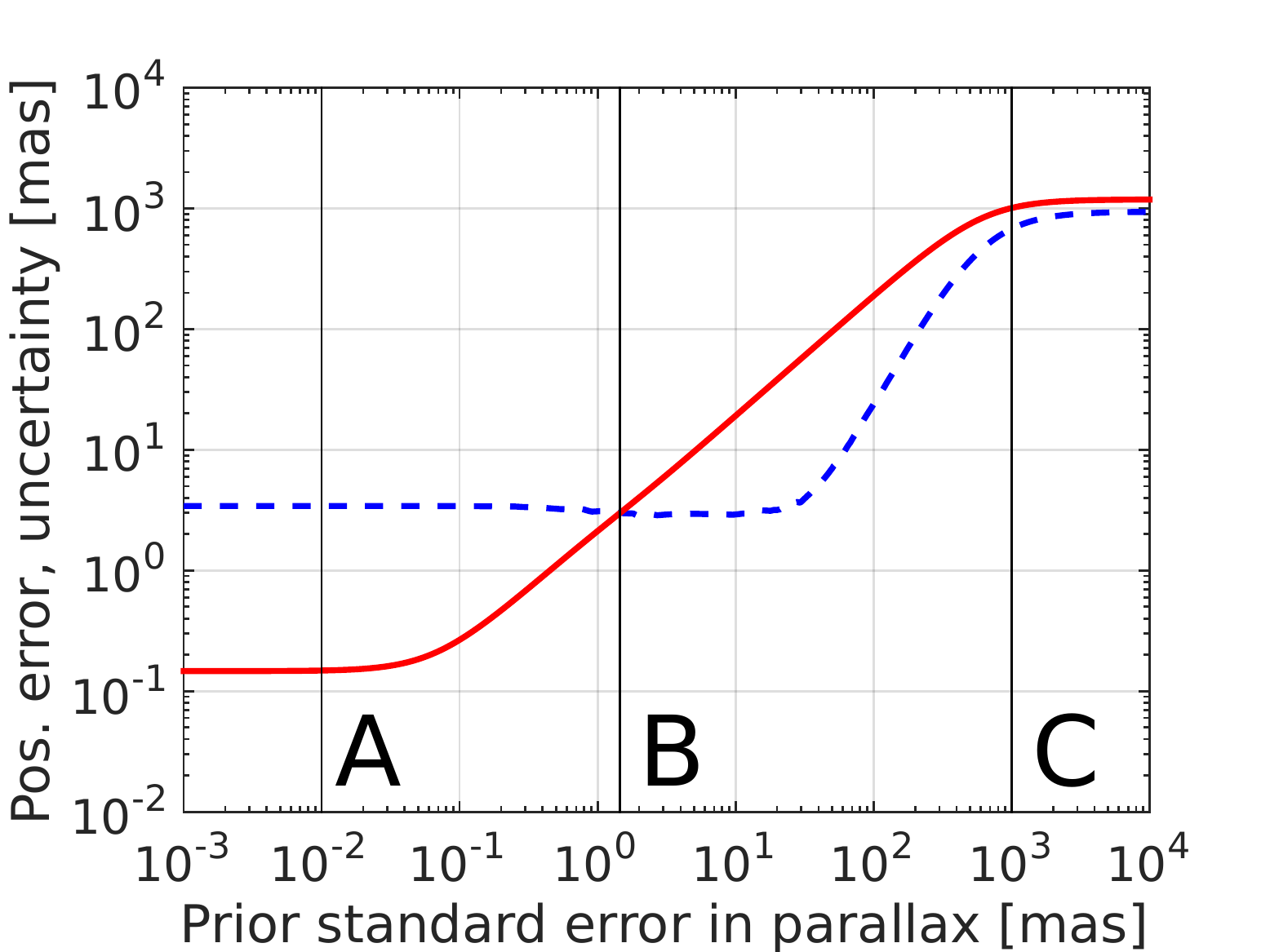}
\caption{Behaviour of the Bayesian position estimate 
as a function of the parallax prior uncertainty $\sigma_{\varpi,p}$, for stars
within one direction and magnitude bin (Table~\ref{tab:GUMS}). 
Blue dashed curve: 90th percentile of the actual position errors. 
Red solid curve: semi-major axis of the 90\% confidence ellipse. The priors labeled A, B, and C refer to the panels in Fig.~\ref{fig:behaviourIndividual}.\label{fig:behaviourSummary}}
\end{figure}
\begin{figure}[t]
\centering
\includegraphics[width=.35\textwidth,clip,trim=0 0 20 0]{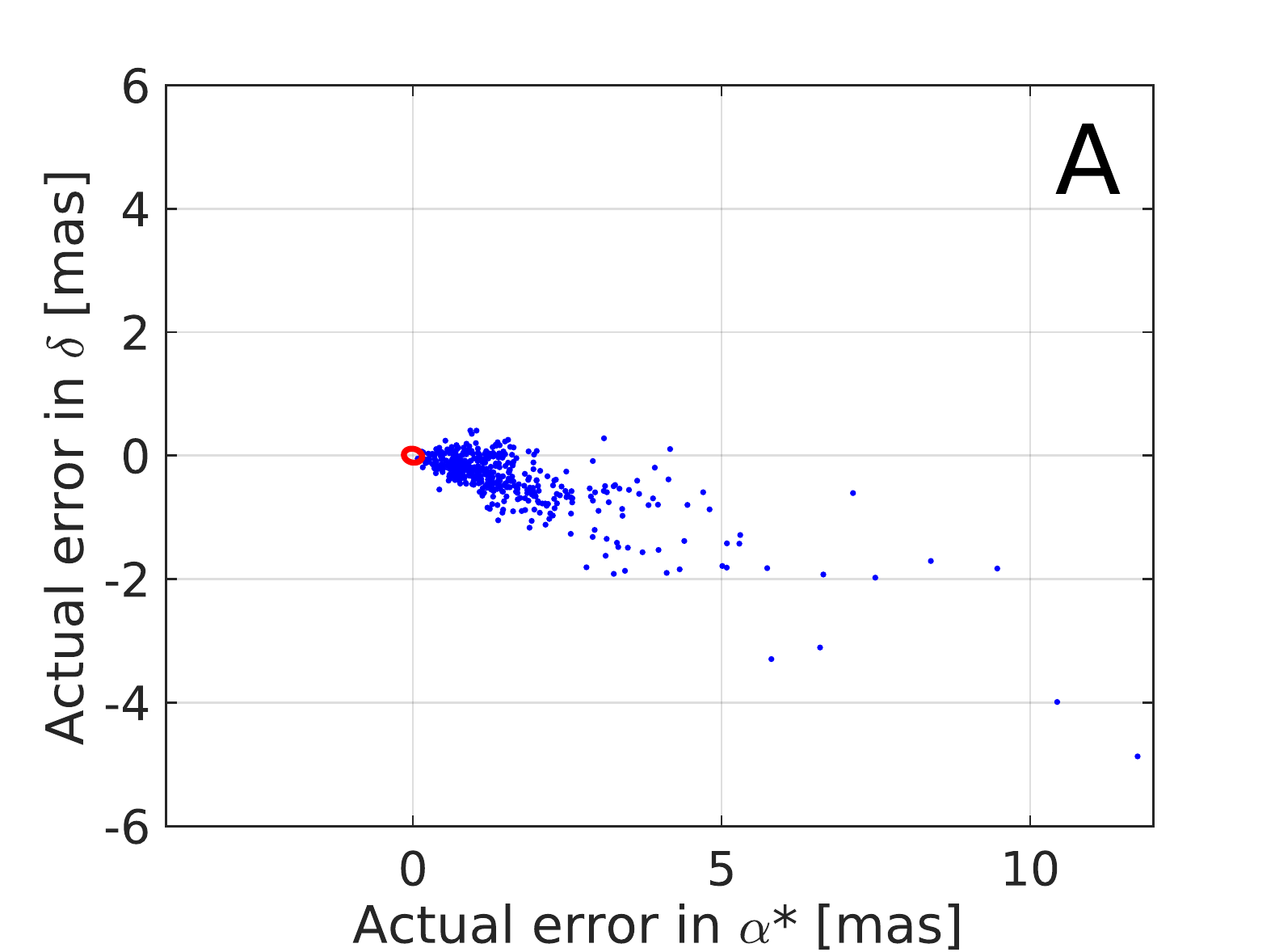}
\includegraphics[width=.35\textwidth,clip,trim=0 0 20 0]{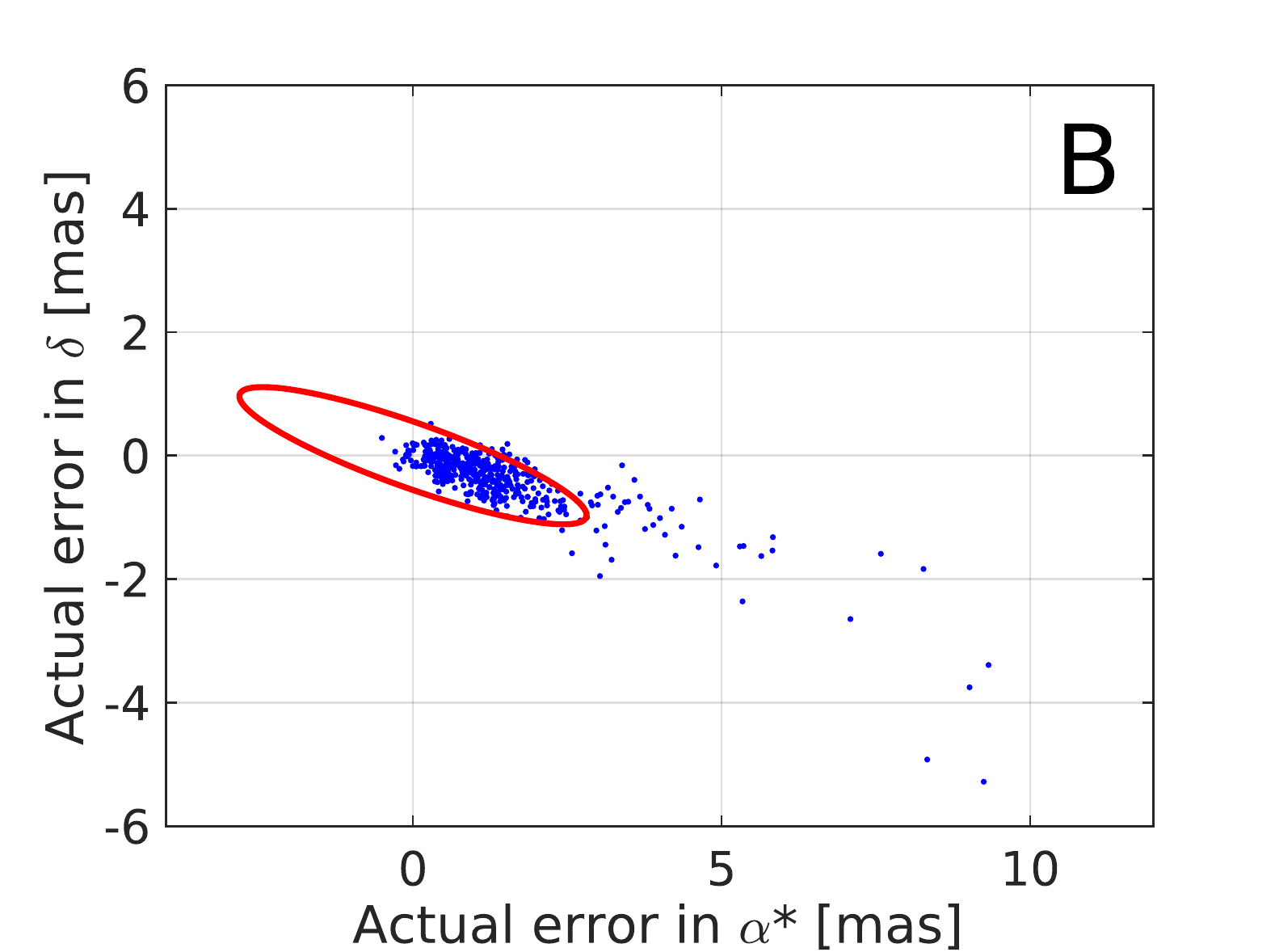}
\includegraphics[width=.35\textwidth,clip,trim=0 0 20 0]{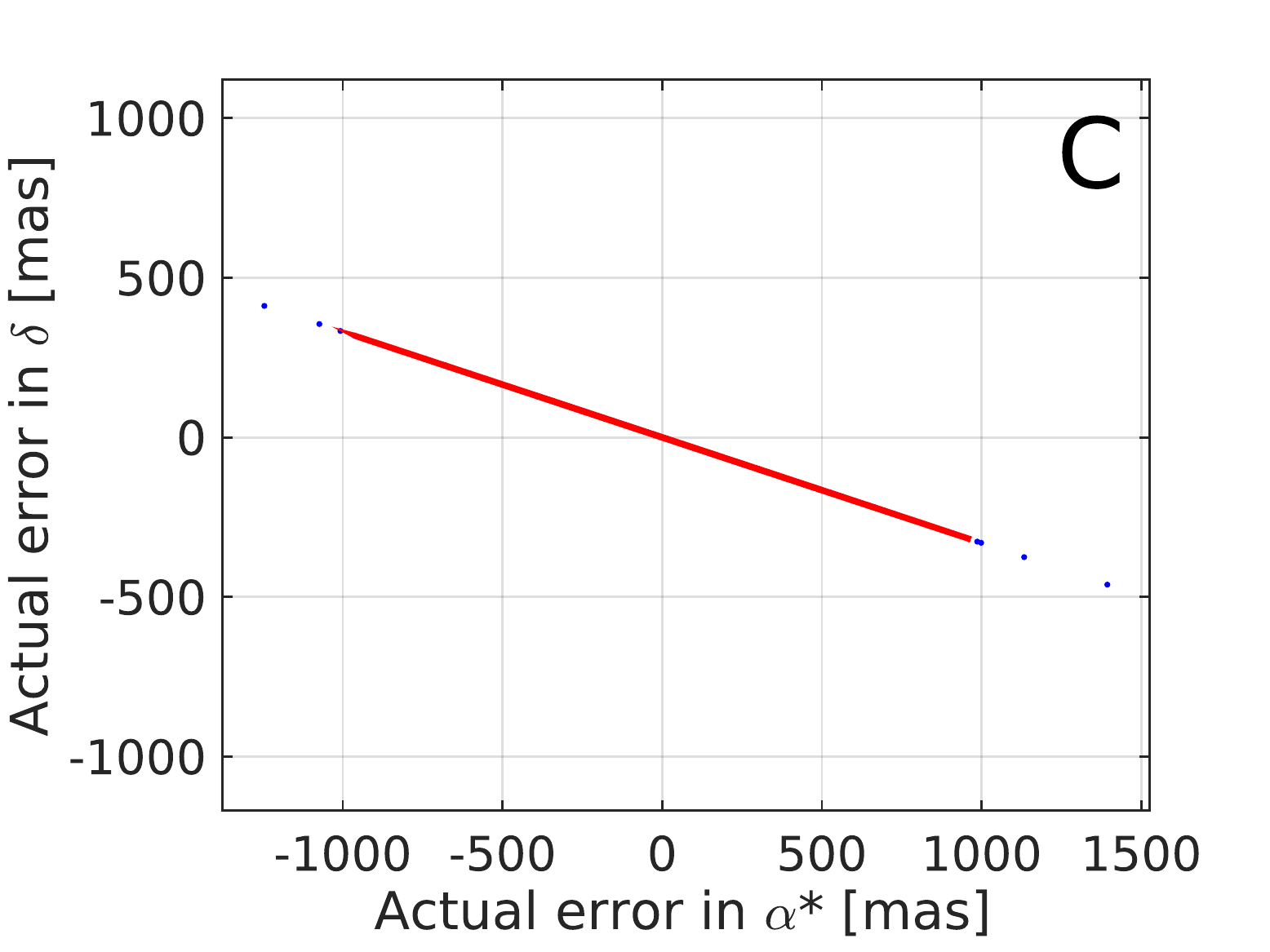}
\caption{Distribution of position errors for the three cases A, B, and C in 
Fig.~\ref{fig:behaviourSummary}.
Blue dots: individual astrometric errors.
Red curve: 90\% confidence ellipse. 
Prior A is too tight and essentially gives a two-parameter solution.
Prior B at the intersection 
of the curves in Fig.~\ref{fig:behaviourSummary}
(semi-major axis of the 90\% confidence ellipse equals the 90th
percentile of the actual errors) produces sensible error estimates.
Prior C is too loose and yields a degenerate solution -- although not
visible in the diagram, 90\% of the points are contained in the extremely
elongated ellipse. 
\label{fig:behaviourIndividual}}
\end{figure}

\begin{table}
\caption{Parameters of the single direction experiments reported in Figs.~\ref{fig:behaviourSummary}, \ref{fig:behaviourIndividual}, and
\ref{fig:othertransits}.\label{tab:GUMS}}
\centering
\small
\begin{tabular}{lccc}
\toprule[\arrayrulewidth]
\toprule[\arrayrulewidth]
\multicolumn{4}{c}{Celestial coordinates}\\
Equatorial:	&$\alpha=157.5\degr$ & $\delta=0.0\degr$\\
Ecliptic:  	&$\lambda=159.2\degr$ &$\beta=-8.8\degr$\\
Galactic:	&$l=245.7\degr$ &$b=+46.5\degr$\\
\midrule[\arrayrulewidth]
\multicolumn{4}{c}{Pencil beam parameters}\\
\multicolumn{2}{l}{Radius of beam:} & 1\degr\\
\multicolumn{2}{l}{Magnitude range:} & $G=15\pm 0.5$~mag\\
\multicolumn{2}{l}{Number of stars in GUMS: } & 458\\
\midrule[\arrayrulewidth]
\multicolumn{4}{c}{Observations according to Gaia's Nominal Scanning Law}\\
Date and time (UTC) & FOV & pos. angle & \# \\
2014-Oct-30 17.0h & P&230\degr & 1\\
2014-Oct-30 18.8h & F&230\degr &1\\
2014-Nov-20 17.0h & P&156\degr &2\\
2014-Nov-20 18.8h & F&156\degr &2\\
2014-Dec-19 16.6h & P&247\degr & 3\\
2014-Dec-19 18.4h & F&247\degr & 3\\
2015-Apr-29 05.2h & P&341\degr & 4\\
2015-Apr-29 07.0h & F&342\degr & 4\\
2015-May-23 18.9h & F&\phantom{0}65\degr & 5\\
2015-Jun-21 04.9h & P&344\degr &6\\
2015-Jun-21 06.7h & F&344\degr & 6\\
2015-Nov-09 06.1h & P&238\degr& 7\\
2015-Nov-09 07.9h & F&238\degr& 7 \\
2015-Dec-29 11.8h & P&243\degr&8\\
2015-Dec-29 13.6h & F&244\degr&8\\
\bottomrule[\arrayrulewidth]
\end{tabular}
\tablefoot{The P and F in the list of observations stand for preceding and following field of view (FOV).
The position angle (third column) is the direction in which the FOV scans across the star, with $0\degr$
towards local North and $90\degr$ towards local East.
The last column (\#) is a sequential numbering of transit groups that are distinct in time and/or direction.}
\end{table}
\section{Prior in an astrometric solution}
\subsection{Framework and basic assumptions\label{sec:framework}}
In order to systematically evaluate the effect of the prior on the astrometric
performance we have developed Matlab scripts which
compute the Bayesian position estimates for a set of simulated stars.
The true stellar parameters are taken from the Gaia Universe Model Snapshot
(GUMS; \citealt{2012A&A...543A.100R}).
Gaia observations are simulated using the Gaia Nominal Scanning Law
\citep{2010IAUS..261..331D} with initial precession and scan phase conditions
consistent with the real mission from October 2014 until the end of 2015.
The astrometric parameters are estimated as described in
Sect.~\ref{sec:theory}. For the initial analysis it is assumed that the spacecraft attitude and
instrument calibration are known, so that the solution only involves the five astrometric
parameters of each star. The posterior covariance and astrometric parameters
are computed and compared for different combinations of magnitude range, position
on the sky, as well as number of observations and their temporal distribution.

We then experiment with varying priors for parallax and proper motion in the
different scenarios. Applying such prior knowledge aids the astrometric solution by constraining parallax and proper motion to small values, without forcing them to be strictly zero. In the present
experiments the prior parallax and proper motion are centred on zero with
Gaussian uncertainties $\sigma_{\varpi, p}$ and  $\sigma_{\mu, p}$, respectively.
The largest known stellar parallax is 768~mas but typical parallaxes are much smaller than that.
$\sigma_{\varpi, p}$ is therefore in the few mas regime.
The proper motion depends on the parallax through the expression for
the transverse space velocity $v_T = A \mu / \varpi$, where $A \simeq
4.74$~km~s$^{-1}$~yr. Linear velocities in the Galaxy are of the order of
30--300~km~s$^{-1}$ and we therefore typically expect $\mu / \varpi \simeq 6$--$60$~yr$^{-1}$. At magnitude 15 the median ratio in GUMS is 10~yr$^{-1}$.
For the ratio $\mathcal{R} = \sigma_{\mu, p} / \sigma_{\varpi, p}$ we have
experimented with values in the range 1--60~yr$^{-1}$ and found the results to be
relatively insensitive to this choice. Using a value of $\mathcal{R} =
10$~yr$^{-1}$ provides reasonable results in all cases, and we adopt this value
in the rest of this paper.

\subsection{Behaviour of the solution as a function of the prior}

For an initial understanding of how the astrometric results depend on the choice of prior
we show a representative example from our experiments.
For one particular position on the sky we took stars from GUMS
of a certain apparent $G$~magnitude in a one degree
pencil beam (Table~\ref{tab:GUMS}). The framework described in Sect.~\ref{sec:framework}
was used to simulate shorter or longer observation intervals of Gaia.
Using one to eight distinct transits (Table~\ref{tab:GUMS}, bottom section), we
obtain the actual errors and formal uncertainties of the resulting position parameters for
each observation interval as a function of prior size $\sigma_{\varpi, p}$.

Figures~\ref{fig:behaviourSummary} and \ref{fig:behaviourIndividual} give the
detailed results for an observation
interval containing two transits that are distinct in time and angle (\#1 and
\#2 in Table~\ref{tab:GUMS}). Figure~\ref{fig:behaviourSummary} summarizes how
the actual errors and formal uncertainties vary as
functions of $\sigma_{\varpi, p}$. The sigmoid shape of the red curve
describing the formal uncertainties is analytically explained in
Appendix~\ref{sec:appendix}. 

Let us first look at the behaviour of the
solution when a very tight prior is applied, e.g.\ $\sigma_{\varpi,p}=0.01$~mas
as indicated by the vertical line at A in Fig.~\ref{fig:behaviourSummary}. The
resulting solution (both with regard to the actual position errors and their
formal uncertainties) is practically
equivalent to solving only for the two position parameters, where the parallax
and proper motion are implicitly assumed to be zero. In this regime the actual errors
(dashed curve) are much larger than the formal position uncertainties (solid curve)
due to the neglected parallax and proper motion. This is
further illustrated by the top panel (prior A) in
Fig.~\ref{fig:behaviourIndividual}, where the 90\% confidence ellipse (red)
only contains a small fraction of the actual errors (blue dots).

Moving from tight to looser priors (increasing $x$-axis values in Fig.~\ref{fig:behaviourSummary}), 
the solution becomes less constrained and the formal uncertainties
necessarily increase. For $\sigma_{\varpi,p} \ge 30$~mas the size of the actual
errors increases, since with two distinct transits the Gaia data alone are
insufficient to determine all five parameters in the solution. Using a very
loose prior, for example prior C illustrated in the bottom panel of Fig.~\ref{fig:behaviourIndividual}, the
astrometric solution becomes almost degenerate, though the formal uncertainties
still correctly describe the actual errors. The intersection point marked with
letter B in Fig.~\ref{fig:behaviourSummary} would be a reasonable compromise, where the solution
is as precise as permitted by the available data, while the formal
uncertainties correctly characterize the actual position errors.  This is
illustrated in the middle panel (prior B) of
Fig.~\ref{fig:behaviourIndividual}, where most of the actual error points lie
within the confidence ellipse.

The actual position errors in panels A and B in
Fig.~\ref{fig:behaviourIndividual} are skewed in a direction depending on the
position of the satellite in its orbit around the Sun. The offset of the error
cloud from the origin depends on the sizes of the parallaxes and proper
motions.

\begin{figure*}[t]
\sidecaption
\begin{minipage}[b]{12cm}
\includegraphics[width=.50\textwidth,trim=0 50 40 20, clip]{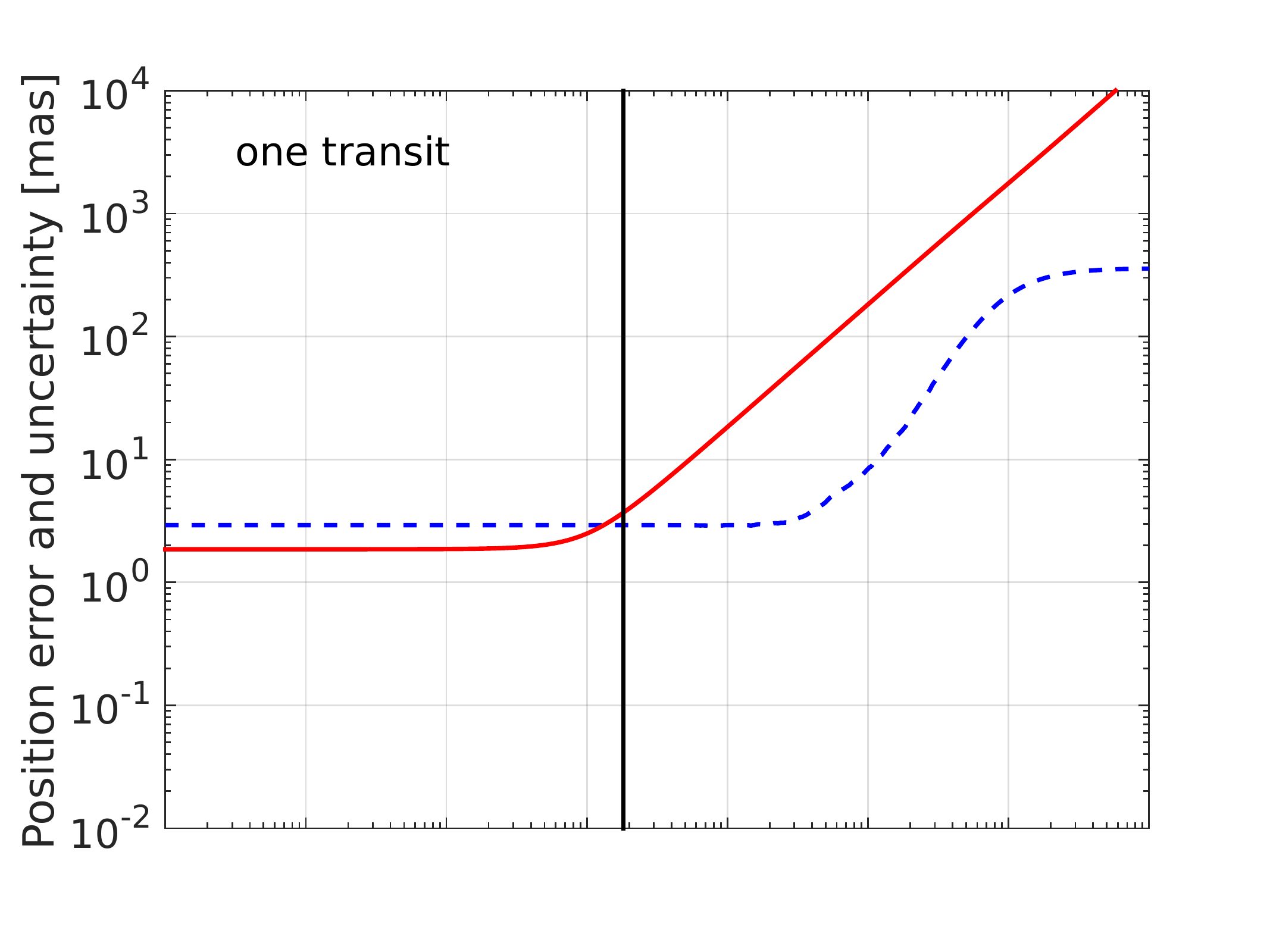}
\includegraphics[width=.50\textwidth,trim=0 50 40 20, clip]{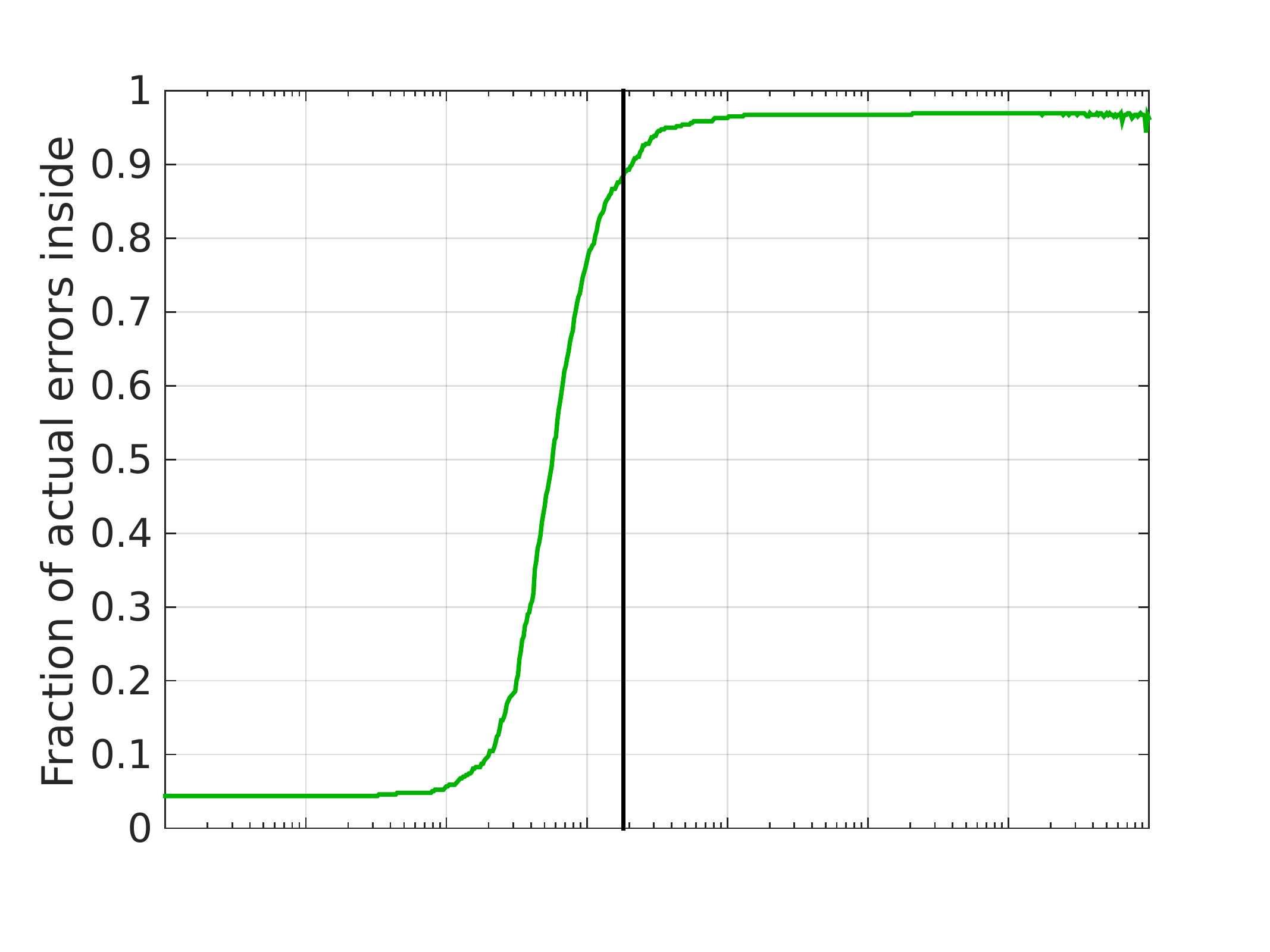}
\includegraphics[width=.50\textwidth,trim=0 50 40 20, clip]{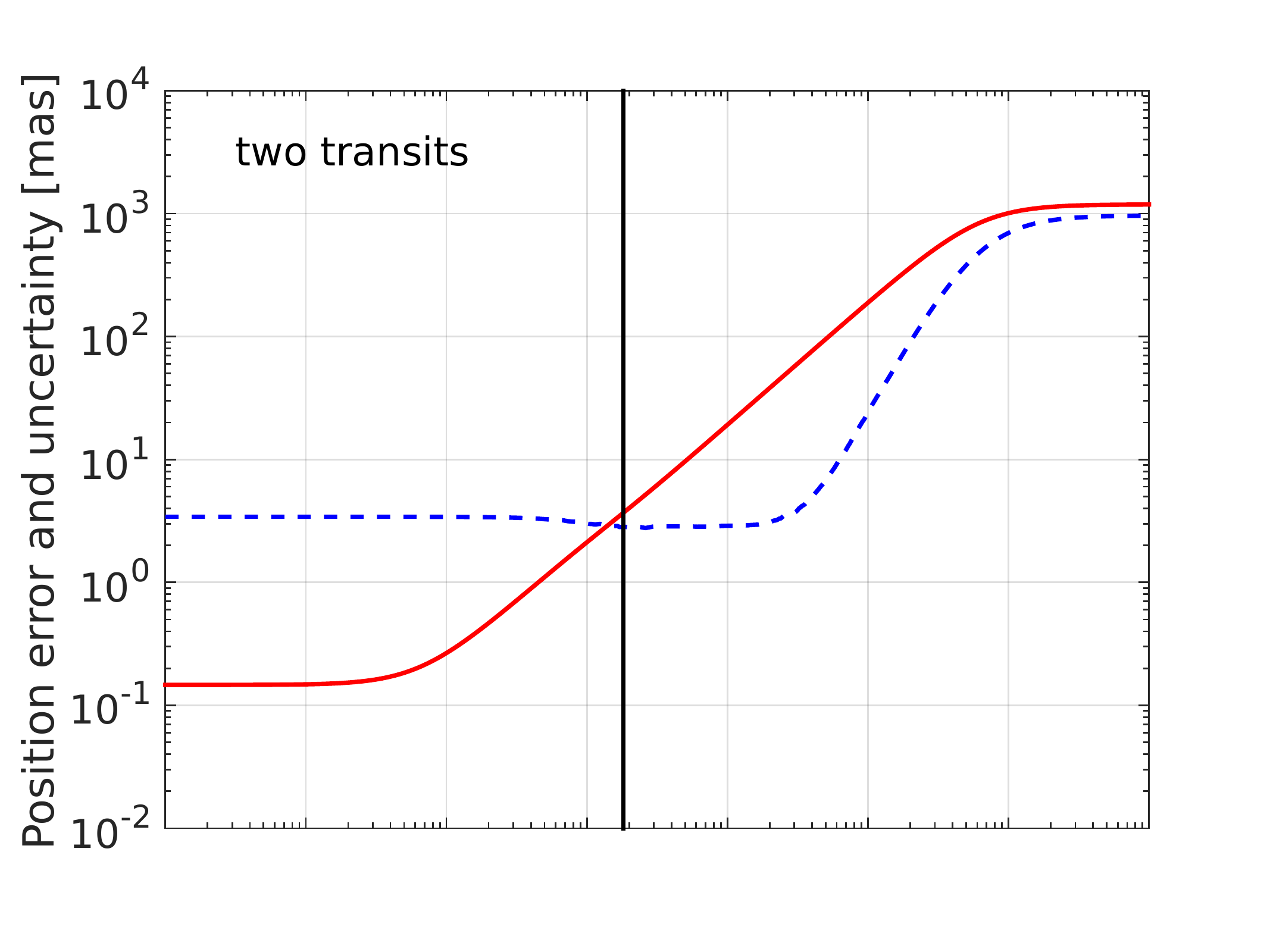}
\includegraphics[width=.50\textwidth,trim=0 50 40 20, clip]{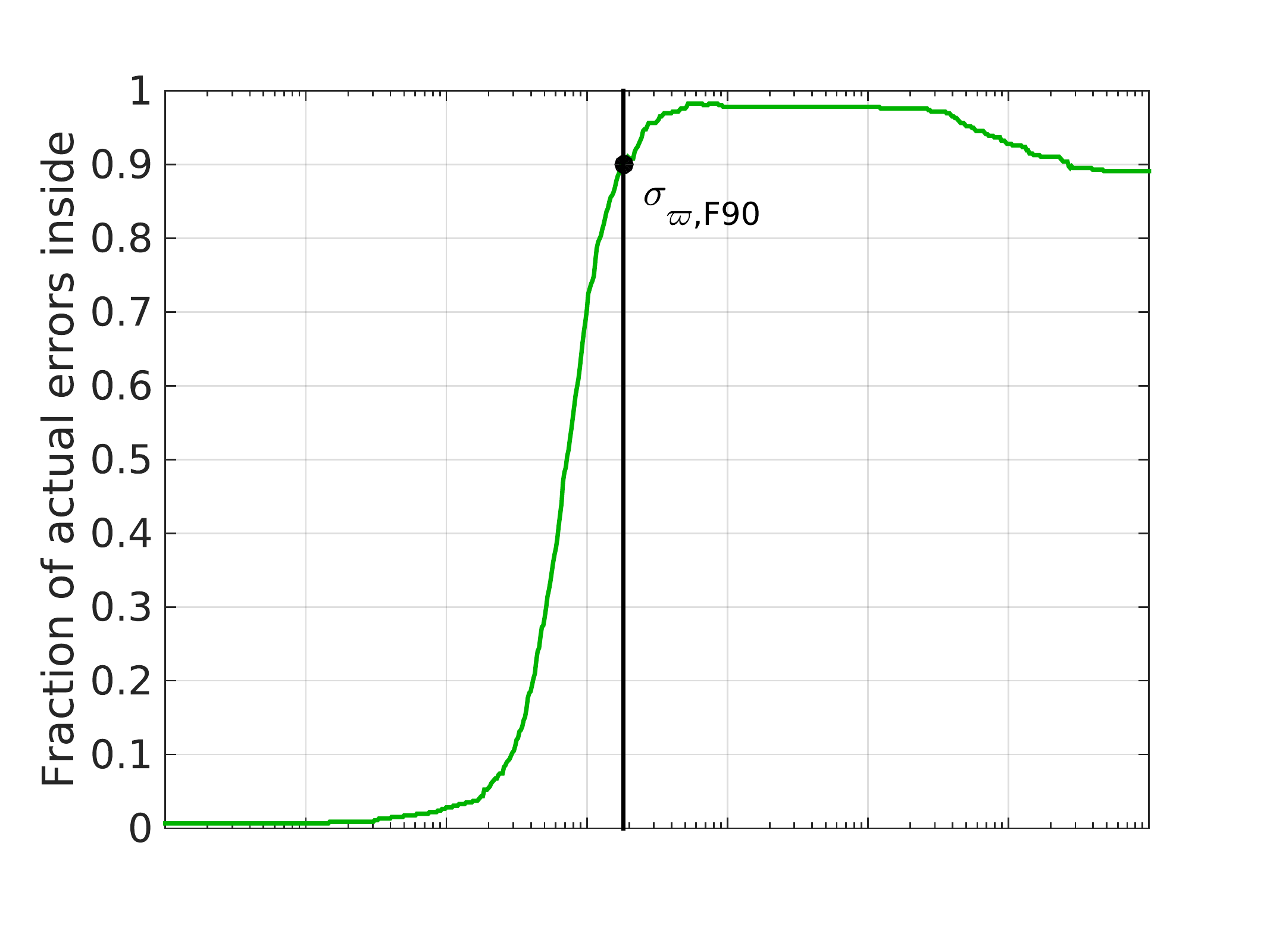}
\includegraphics[width=.50\textwidth,trim=0 50 40 20, clip]{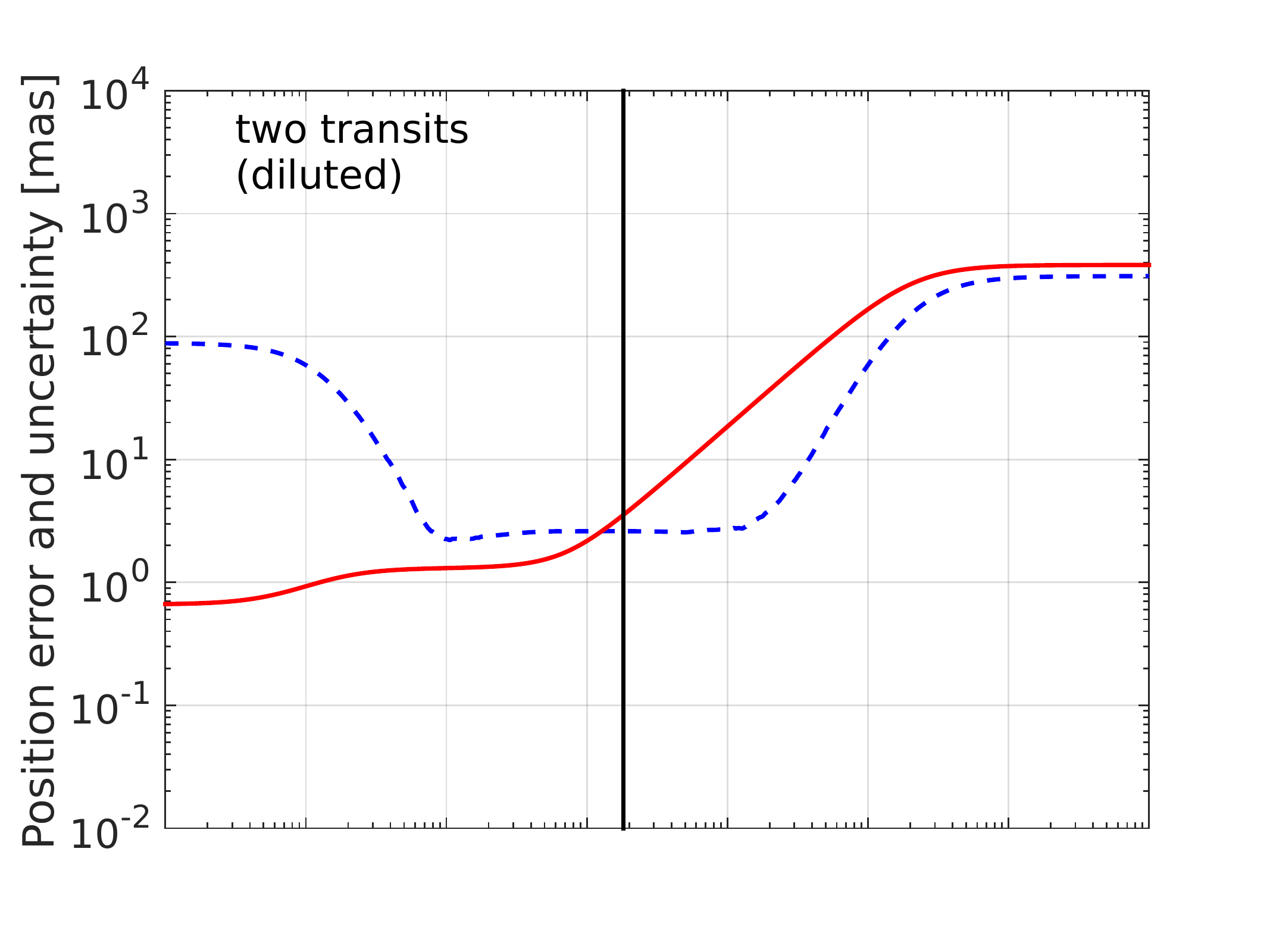}
\includegraphics[width=.50\textwidth,trim=0 50 40 20, clip]{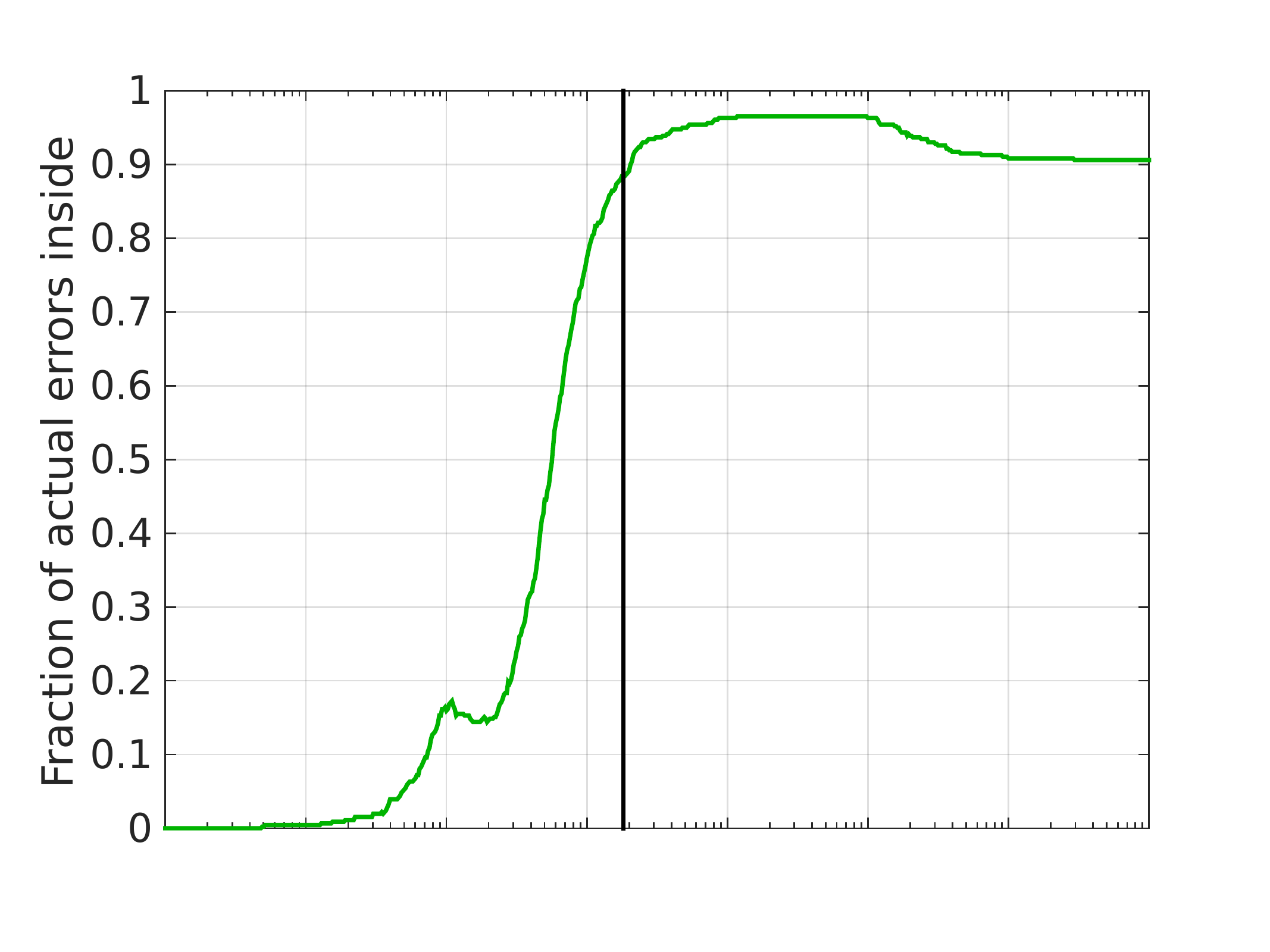}
\includegraphics[width=.50\textwidth,trim=0 50 40 20, clip]{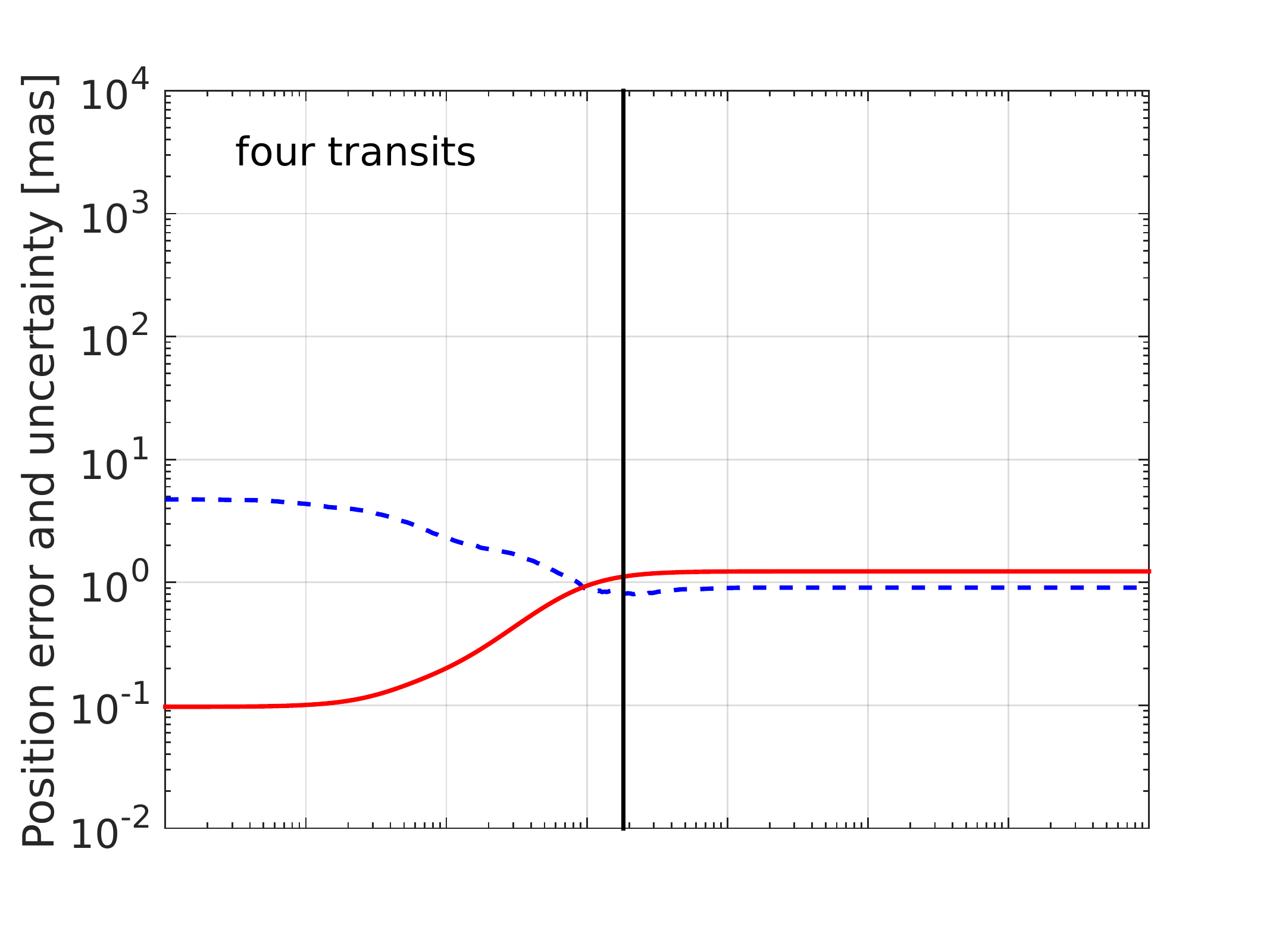}
\includegraphics[width=.50\textwidth,trim=0 50 40 20, clip]{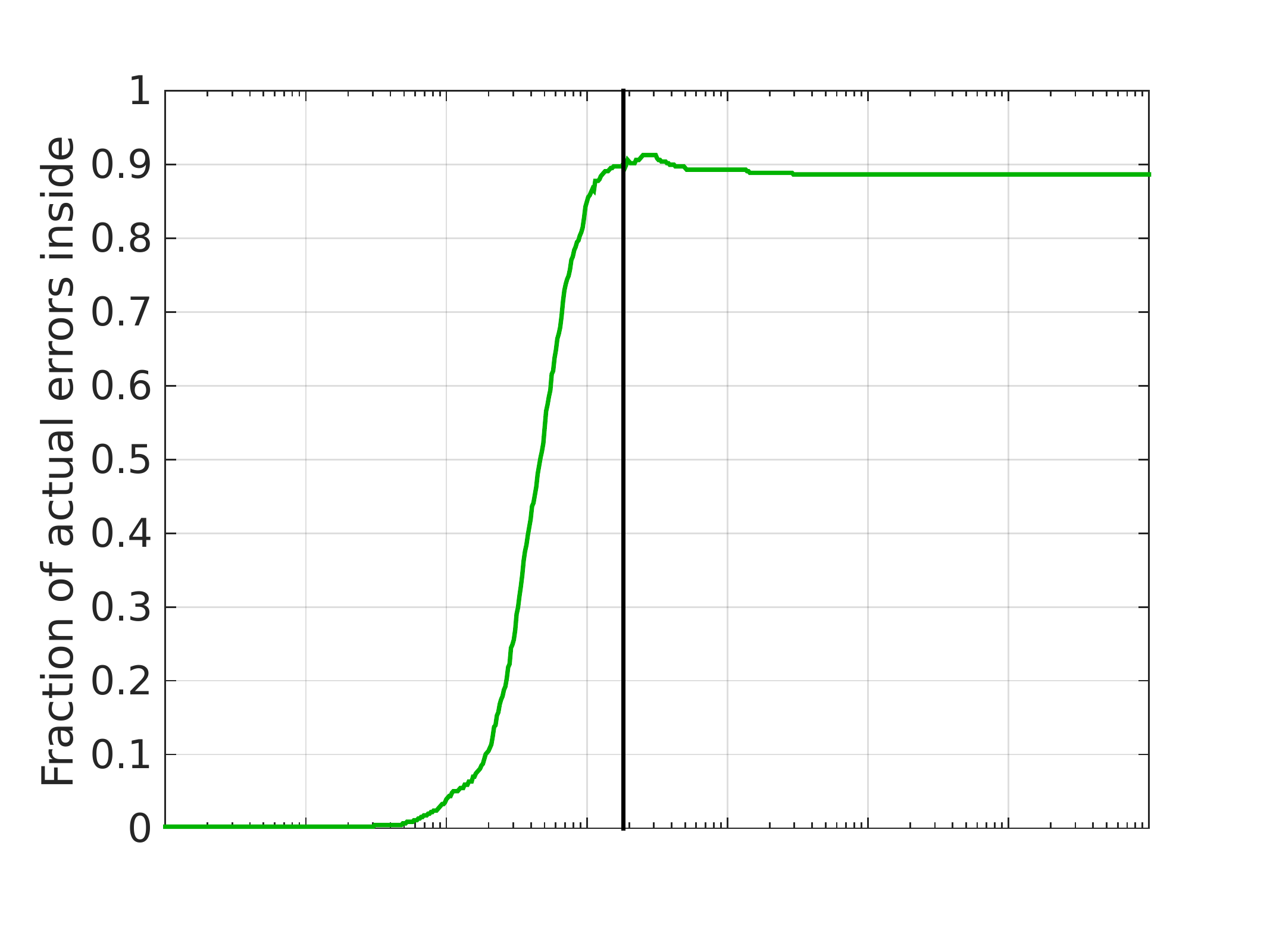}
\includegraphics[width=.50\textwidth,trim=0 0 40 20, clip]{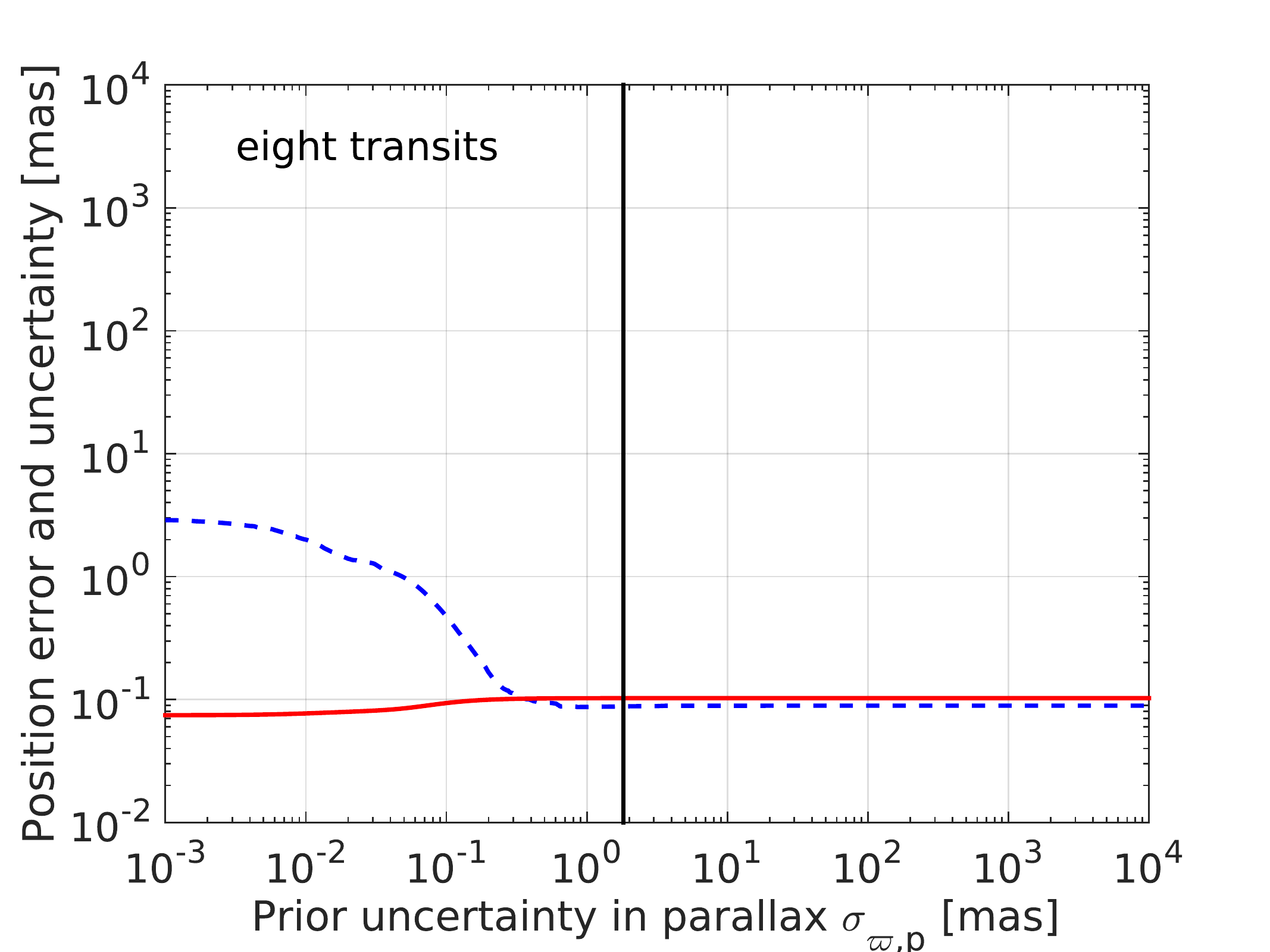}
\includegraphics[width=.50\textwidth,trim=0 0 40 20, clip]{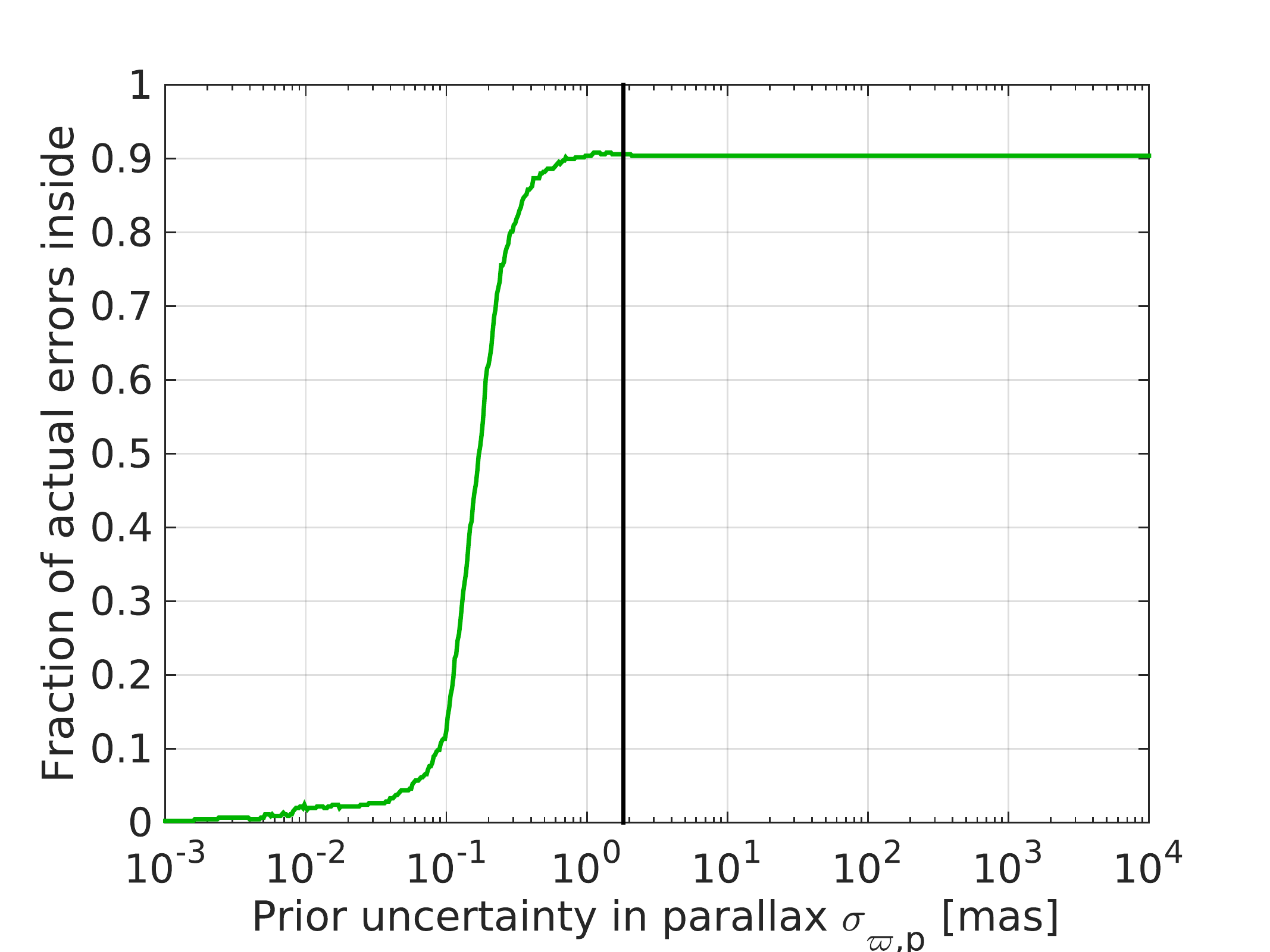}
\end{minipage}
\caption{Behaviour of the position error and uncertainty with varying priors, for stars in
the direction and magnitude bin specified in Table~\ref{tab:GUMS}.
The rows display the behaviour for different numbers of distinct transits: one,
two, two (diluted), four, and eight distinct transits. The diluted case uses
the first and last transit from Table~\ref{tab:GUMS} instead of the first two
transits.
Left column: size of actual position errors (blue dashed) and their formal uncertainties (red solid), for stars in the same
direction and magnitude bin, as a function of the prior uncertainty. Right column: fraction of actual
errors contained by the formal error ellipse. \label{fig:othertransits}
The optimum prior
$\sigma_{\varpi,F90}$ is chosen based on the observation interval containing
two distinct consecutive transits (second row). This prior is replicated in all
other panels.}
\end{figure*}

\subsection{Criterion for the optimum prior uncertainties\label{sec:choice}}

The two quantities represented by the dashed and solid curves in Fig.~\ref{fig:behaviourSummary} are not exactly comparable:
one is the radius of the circle, centred on zero, that contains 90\% of the
actual errors; the other is the semi-major axis of the confidence ellipse.
Using their point of intersection to optimise the prior, as suggested in
the previous section, is therefore slightly illogical. We have instead adopted
a different and much simpler criterion based on the confidence ellipse:
the optimum prior should be such that 90\% of the actual position errors
are contained by the 90\% confidence ellipse as calculated from the covariance
matrix. The smallest $\sigma_{\varpi,p}$ fulfilling this condition is in the following called
$\sigma_{\varpi,F90}$ and is illustrated in the second row of Fig.~\ref{fig:othertransits}.
The left diagram replicates the curves for two distinct transits previously
shown in Fig.~\ref{fig:behaviourSummary}. The right diagram shows
the corresponding fraction of actual errors contained in the 90\% confidence
ellipse. The prior choice $\sigma_{\varpi,F90}$, marked by the solid vertical
line and replicated in all panels of the figure, is in fact quite close to the
intersection of the two curves in the left diagram.

So far we have limited our discussion to a scenario with two distinct transits. This
is the case where the prior information is expected to be most critical: two distinct
along-scan observations may suffice to determine a sensible position, but are
always insufficient for a full five-parameter solution; on the other hand, three
distinct transits in principle already allow a five parameter solution if both along-
and across-scan information is used.
We adopt $\sigma_{\varpi,F90}$ based on the two-transit case and use it also
in other scenarios with more or less observations. That the same prior works
in these cases has been verified through simulations. Examples are given in
Fig.~\ref{fig:othertransits}, where the different rows show the behaviour
of the astrometric solution for observation intervals containing one, two, four,
and eight distinct transits, including one diluted case of two transits
separated by 14~months. The prior $\sigma_{\varpi,F90}$ determined from the two-transit
scenario, and indicated by the solid vertical line in all panels, yields in all
cases a solution where the size of the actual position errors (as measured by the
90th percentile) is close to its minimum, together with a realistic 90\% confidence
ellipse.

It is also evident that $\sigma_{\varpi,F90}$ is a lower limit for a
suitable prior. Increasing $\sigma_{\varpi,p}$ by up to a factor $\sim$10 keeps the actual
position errors at the same level while providing the same or a more conservative formal
uncertainty estimate, whereas using a smaller prior would underestimate the errors.
In Appendix~\ref{sec:appendix} we briefly address the effect of the prior
uncertainty on the posterior error estimate from an analytical point of view.

\begin{figure*}
\includegraphics[width=0.33\textwidth,clip,trim=30 10 50 20]{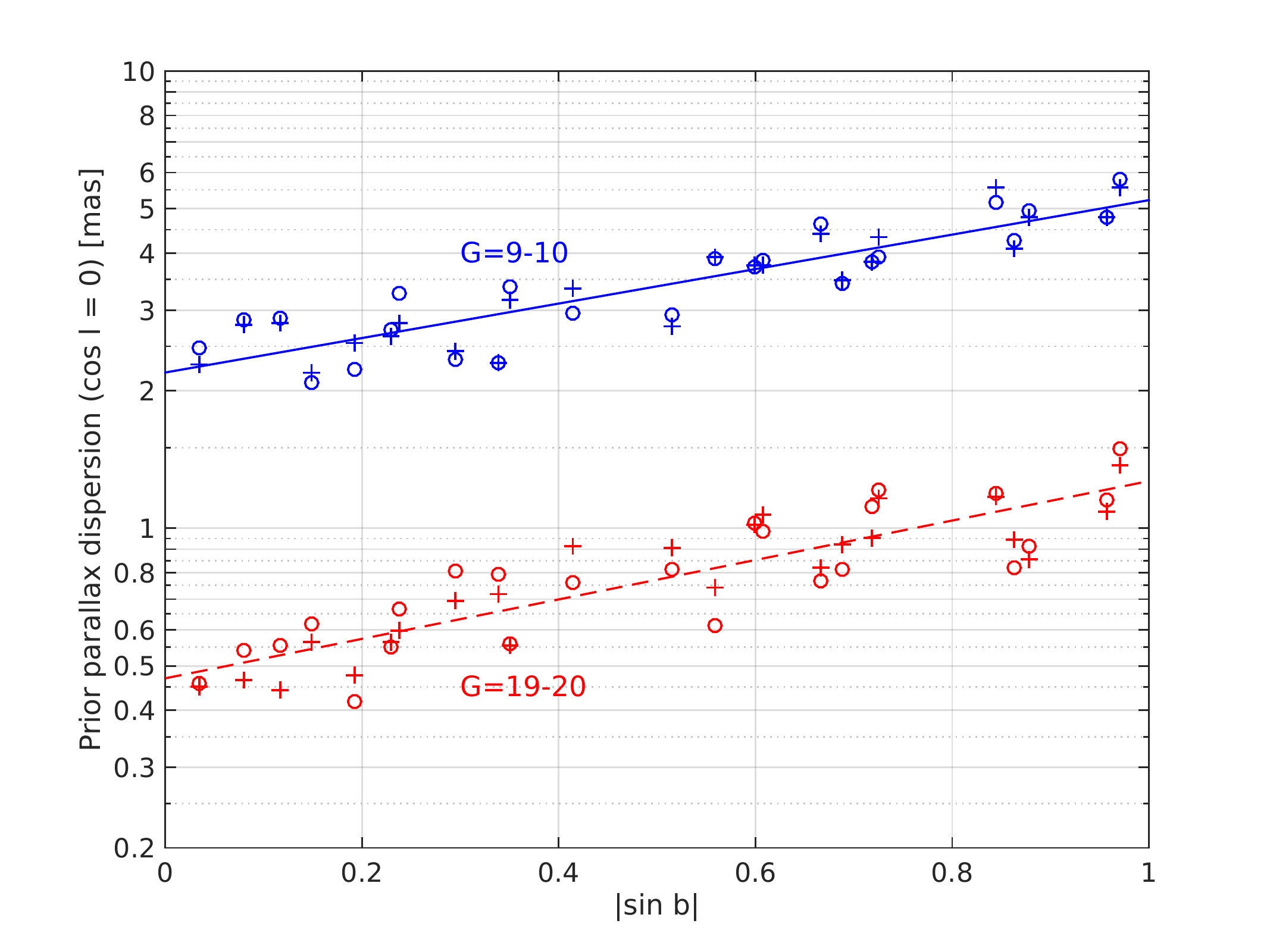}
\includegraphics[width=0.33\textwidth,clip,trim=30 10 50 20]{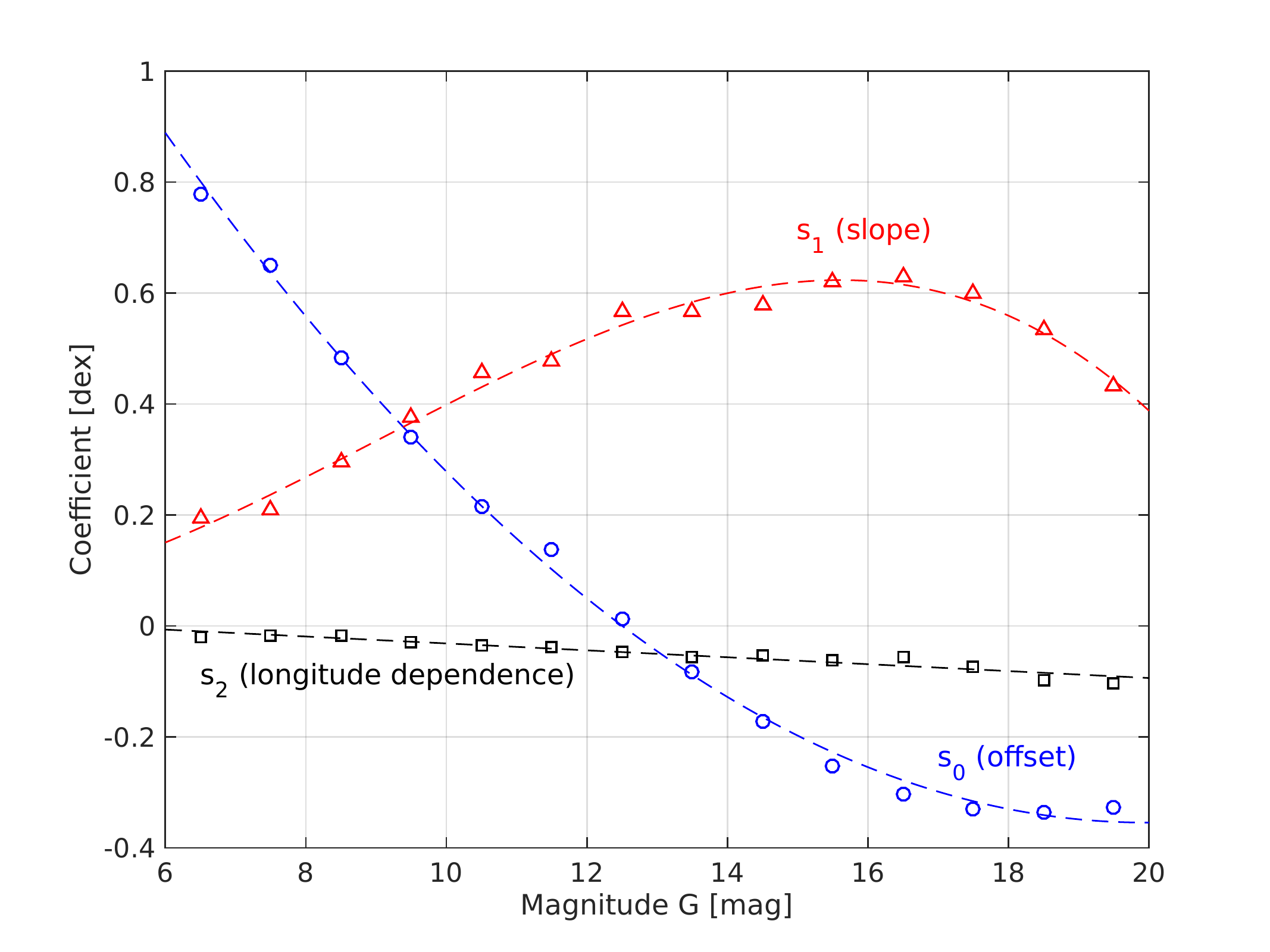}
\includegraphics[width=0.33\textwidth,clip,trim=30 10 50 20]{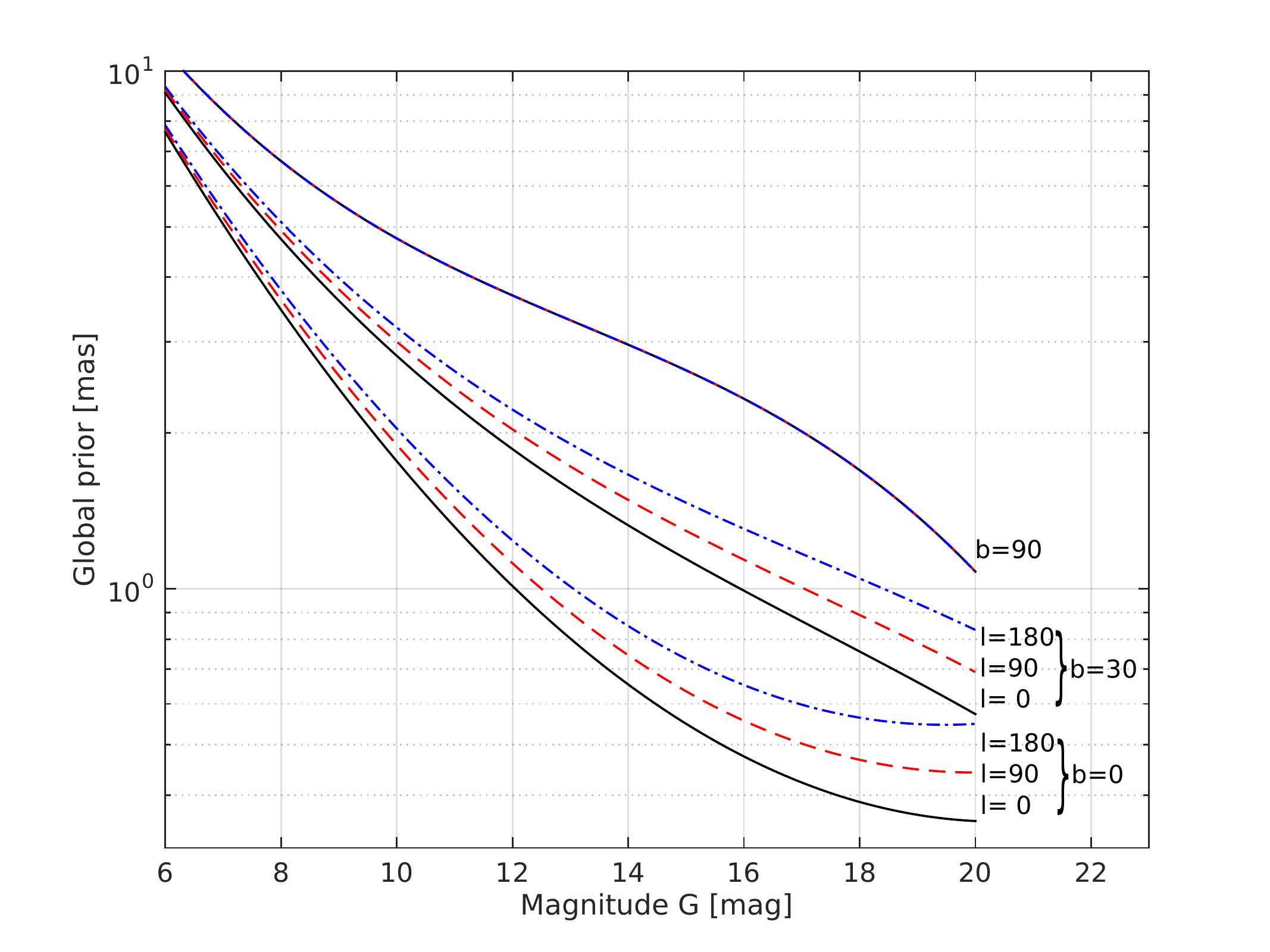}
\caption{Illustrations of the all-sky approximation to the parallax prior $\sigma_{\varpi,F90}$.
Left: Representative example of the linear fit in Eq.~(\ref{eq:linearfit}) of the logarithm of the parallax prior
$\sigma_{\varpi,F90}$ as a function of galactic latitude $b$, for two magnitude
ranges $G=9$--$10$ (blue), and $G=19$--$20$ (red).
The small dependence on galactic longitude (the third term in
Eq.~\ref{eq:linearfit}) has been subtracted leaving no systematic dependence on
galactic longitude $l$, as shown by the different symbols ($\circ$ for $\cos l >0$, + for
$\cos l <0$).\label{fig:linearfit}
Middle: Variations of the coefficients $s_0$, $s_1$, and $s_2$ with $G$ magnitude, and polynomial fits according to Eqs.~(\ref{eq:fitf0})--(\ref{eq:fitf2}). Right: The prior $\sigma_{\varpi,F90}$ in Eq.~(\ref{eq:linearfit}) as a function of magnitude,
for different latitudes (lowest to highest group: $b=0\degr$, 30\degr, and 90\degr) and different longitudes
(black solid $l= 0\degr$, red dashed 90\degr, blue dotted-dashed 180\degr). \label{fig:abc}\label{fig:priorsize}}
\end{figure*}

\subsection{Prior uncertainty as function of magnitude and direction\label{sec:allsky}}

In Sect.~\ref{sec:choice} we described the determination of an
optimum value of $\sigma_{\varpi,p}$, called $\sigma_{\varpi,F90}$, for one particular
direction and magnitude interval.
We repeated this experiment for different directions and magnitude bins
($G=6$--20, in steps of 1~mag). We find that 48 directions uniformly
distributed on the sky are sufficient to sample the large-scale structures of the
underlying Galaxy model. As expected, $\sigma_{\varpi,F90}$ is a strong function of magnitude (fainter stars
being on average more distant), and to a lesser extent dependent on direction
(because of extinction and the spatial distribution of stars in our Galaxy).%
\footnote{Statistically, $\sigma_{\varpi,F90}$ is closely related to the
distribution of parallaxes in GUMS. Inspection of the distribution shows that,
for any given magnitude and direction, it is roughly equal to 0.5--0.8 times
the 90th percentile of the parallaxes.}
For a given magnitude bin we find that a reasonable fit to the individual data
points is provided by
\begin{equation}
\log_{10}\sigma_{\varpi,F90} = s_0 + s_1 |\sin b| + s_2 \cos b \cos l \, \label{eq:linearfit}
\end{equation}
(see the left panel of Fig.~\ref{fig:linearfit}).
The variations of the coefficients $s_0$, $s_1$, and $s_2$ with $G$ magnitude
bins are shown in the middle panel of Fig.~\ref{fig:abc}. They are well
approximated by simple polynomials in~$G$:
\begin{align}
s_0(G)& = 2.187 - 0.2547 G + 0.006382 G^2\label{eq:fitf0}\\
s_1(G)& =  0.114 - 0.0579 G +  0.01369 G^2- 0.000506 G^3 \label{eq:fitf1}\\
s_2(G)& =  0.031 - 0.0062 G\label{eq:fitf2}
\end{align}
The size of the fitted $\sigma_{\varpi,F90}$ prior is illustrated in the right
panel of Fig.~\ref{fig:priorsize}.
For the astrometric solution of an arbitrary
star of magnitude $G$ at galactic coordinates $(l,b)$ the prior normal matrix
to be used in Eqs.~(\ref{eq:priorincoporation})--(\ref{eq:priorincoporation2})
is then
\begin{equation}
\label{eq:prior}
\vec{N}_{p} = \textrm{diag}(0,\ 0,\
\sigma_{\varpi,F90}^{-2},\ \sigma_{\mu,F90}^{-2},\ \sigma_{\mu,F90}^{-2})\, ,
\end{equation}
where $\sigma_{\varpi,F90}(l,b,G)$ is given by
Eqs.~(\ref{eq:linearfit})--(\ref{eq:fitf2}), and $\sigma_{\mu,F90} =
\mathcal{R} \sigma_{\varpi,F90}$, where $\mathcal{R} = 10$~yr$^{-1}$.

An extension of $\sigma_{\varpi,F90}(l,b,G)$ to
fainter stars is non-trivial, since GUMS is only complete to $G=20$. For fainter
stars the value at $G=20$ should be used since it provides a conservative
(over)estimate. For stars brighter than $G=6$ it might be preferable to make
solutions directly using priors from the Hipparcos and Tycho-2 catalogues.

In principle more sophisticated priors could be considered, which take into
account photometric, spectroscopic, or other auxiliary information. For
example, blue stars have on average smaller parallaxes than red stars, and
for identified extra-galactic objects the prior uncertainty could be much smaller.
However, such information may be unavailable precisely in the cases where a
prior is needed. On the other hand, a direction and an approximate magnitude
are always available, and allow us to define a general prior.

\begin{table*}
\caption{Simulation results for stars with
95\% diluted five year observation histories. The spacecraft attitude was
determined by a separate solution from well-observed stars. A prior uncertainty $\ge \sigma_{\varpi, F90}$ provides a sensible solution for all stars.
\label{tab:results}}
\centering
\small
\begin{tabular}{crrrrrr}
\toprule[\arrayrulewidth] \toprule[\arrayrulewidth]
Prior $\sigma_{\varpi,p}$ & \multicolumn{3}{c}{Fraction in 90\% conf. ellipse} & \multicolumn{3}{c}{Actual position errors [mas]} \\
\cmidrule[0.2pt](lr){2-4}
\cmidrule[0.2pt](lr){5-7}
& $G\simeq 11$	& $G\simeq 15$	& $G\simeq 19$ & $G\simeq 11$	& $G\simeq 15$	& $G\simeq 19$ \\
\midrule[\arrayrulewidth]  
\multicolumn{7}{c}{Subset of stars with $\le$4 field-of-view transits}\\
\midrule[\arrayrulewidth]  
none (2 parameters)	  	& 0.5\% 	& 1.8\% 	& 13.5\% 	& 33.0 	& 16.3 	& 15.2 \\
$0.01$ mas 		  	& 1.5\% 	& 3.5\% 	& 14.3\% 	& 21.8 	& 12.1 	& 14.8 \\
$\sigma_{\varpi,F90}$		& 90.1\% 	& 91.4\% 	& 91.2\% 	& 7.6 	& 4.3 	& 7.6 \\
$10 \sigma_{\varpi,F90}$    	& 92.7\% 	& 93.3\% 	& 94.4\% 	& 8.4 	& 5.2 	& 10.5\\
$1000$ mas 			& 92.5\% 	& 93.0\% 	& 93.3\% 	& 8.6 	& 7.4 	& 15.5\\
\midrule[\arrayrulewidth]  
\multicolumn{7}{c}{Subset of stars with >4 field-of-view transits}\\
\midrule[\arrayrulewidth]  
none (2 parameters)	  	& 0.3\% 	& 0.8\% 	& 8.6\% 	& 21.0 	& 11.3 	& 9.7 \\
$0.01$ mas 		  	& 3.1\% 	& 5.4\% 	& 10.4\% 	& 6.7 	& 5.0 	& 8.9 \\
$\sigma_{\varpi,F90}$		& 89.4\% 	& 89.9\% 	& 90.3\% 	& 0.2 	& 0.3 	& 1.6 \\
$10 \sigma_{\varpi,F90}$  	& 89.5\% 	& 89.8\% 	& 90.5\% 	& 0.2 	& 0.3 	& 2.0 \\
$1000$ mas 			& 89.5\% 	& 89.8\% 	& 90.0\% 	& 0.2 	& 0.3 	& 2.2 \\
\bottomrule[\arrayrulewidth]  
\end{tabular}
\tablefoot{Column~1: prior uncertainty used in the solution. Columns~2--4: fractions of actual position errors contained in the 90\% confidence ellipses
calculated from the formal covariances; ideally, these values should be around 90\%. Columns~5--7: 90th percentile values of the actual position errors
(estimated minus true value) in mas; these should be as small as possible. For comparison: two parameter solution (position only) without a prior.  }
\end{table*}

\begin{table*}
\caption{Global astrometric solutions using 12~months of simulated
Gaia data. The attitude is determined as part of the solution. Here the prior
uncertainty needs to be relaxed to $10 \sigma_{\varpi, F90}$ to obtain a
sensible solution. Columns as in Table~\ref{tab:results}.
\label{tab:results1yr}}
\centering
\small
\begin{tabular}{crrrrrr}
\toprule[\arrayrulewidth] \toprule[\arrayrulewidth]
Prior $\sigma_{\varpi,p}$ & \multicolumn{3}{c}{Fraction in 90\% conf. ellipse} & \multicolumn{3}{c}{Actual position errors [mas]} \\
\cmidrule[0.2pt](lr){2-4}
\cmidrule[0.2pt](lr){5-7}
& $G\simeq 11$	& $G\simeq 15$	& $G\simeq 19$ & $G\simeq 11$	& $G\simeq 15$	& $G\simeq 19$ \\
\midrule[\arrayrulewidth]  
\multicolumn{7}{c}{Subset of stars with $\le$4 field-of-view transits}\\
\midrule[\arrayrulewidth]  
$\sigma_{\varpi,F90}$	& 76.5\% 	& 86.8\% 	& 92.4\% 	& 1.2 	& 0.7 	& 2.0 \\
$10 \sigma_{\varpi,F90}$ 	& 89.1\% 	& 90.4\% 	& 94.3\% 	& 1.4 	& 1.1 	& 4.2 \\
$1000$ mas				& 89.7\% 	& 89.8\% 	& 90.6\% 	& 3.8 	& 3.4 	& 17.7 \\
\midrule[\arrayrulewidth]  
\multicolumn{7}{c}{Subset of stars with >4 field-of-view transits}\\
\midrule[\arrayrulewidth]  
$\sigma_{\varpi,F90}$			& 32.3\% 	& 53.3\% 	& 88.0\% 	& 0.2 	& 0.2 	& 0.8 \\
$10 \sigma_{\varpi,F90}$  	& 88.6\% 	& 89.3\% 	& 90.4\% 	& 0.1 	& 0.2 	& 1.0 \\
$1000$ mas			& 88.7\% 	& 89.4\% 	& 90.0\% 	& 0.1 	& 0.2 	& 1.1 \\
\bottomrule[\arrayrulewidth]  
\end{tabular}
\end{table*}

\section{Simulation of potential application scenarios\label{sec:globalsolutions}}
In this section we demonstrate the feasibility of the proposed method based on
simulations of potential applications.
We used GUMS to provide simulated `true' parameters for a large catalogue of stars
of different magnitude classes, where we include the $5\times 10^5$ brightest stars
fainter than each of the magnitudes $G=11$, 15, and 19, respectively.
 The software package
AGISLab \citep{2012A&A...543A..15H,2012A&A...538A..77B} was used to simulate Gaia
observations and to perform a global astrometric solution.

In Sect.~\ref{sec:introduction} we described three situations where the
Bayesian approach might be useful: transient objects, faints stars at the
detection limit, and the
processing of short stretches of Gaia data. The
first two situations are similar to the diluted case presented in
Sect.~\ref{sec:choice}. When solving the astrometric parameters we can assume that
an accurate
satellite attitude is known from a previous solution of well-observed stars. The
third situation applies to the first release of Gaia data, where the spacecraft attitude must be
obtained together with the astrometric parameters from the same (insufficient)
data, and therefore is much less accurate than in the previous scenario. We
therefore performed two distinct sets of simulations described hereafter.

\subsection{Stars with very diluted observation histories\label{sec:secondary}}
Here we use an attitude determined by a five year
solution of simulated Gaia data without prior. We then compute the astrometric parameters without
changes to the attitude (a so-called secondary solution) for stars with a highly
diluted observation history. The dilution is simulated by assigning each field
of view transit a 95\% probability of being removed from the solution.  The
average number of retained transits per star is $\simeq 4.4$.

We made a solution using the optimum prior according to Sect.~\ref{sec:allsky}.
Additionally we experimented with a very tight prior
($\sigma_{\varpi,p}=0.01$~mas, analogous to case A in Fig.~\ref{fig:behaviourIndividual})
and a very loose prior ($\sigma_{\varpi,p}=1000$~mas, analogous to case C in
Fig.~\ref{fig:behaviourIndividual}) to check the behaviour of the solution in these
extreme cases. For comparison we also made two runs without any prior, one in which only
the two position parameters were determined, and one with all five astrometric
parameters. In the latter case it was not possible to determine a unique
astrometric solution for all stars, as explained at the end of Sect.~\ref{sec:leastsquare}.

Table~\ref{tab:results} summarizes our results. Only the results for the position
estimates are shown. When using a very tight prior
the results are very similar to a two parameter solution. The
formal uncertainties computed in this solution grossly underestimate the actual
errors. Using the optimum prior $\sigma_{\varpi,F90}$ (or ten times its value) instead yields sensible
estimates of the uncertainties. The use of this prior
not only provides improved uncertainties, but also reduces the actual errors
compared with a two-parameter solution (or a very tight prior). This somewhat
surprising behaviour can be understood from the three bottom left panels in
Fig.~\ref{fig:othertransits}: using a non-zero prior uncertainty provides
the necessary freedom for the solution to accommodate non-zero parallaxes and
proper motions and hence to reduce the actual position errors.

With a very loose prior of
one arc-second the Bayesian approach still results in a numerically stable
astrometric solution (which is not true for solutions without any prior),
including realistic estimates of the positional uncertainties. However the
actual errors are up to a factor two larger than when using a well-chosen prior.

\subsection{The first data release\label{sec:primary}}
Another possible application of the proposed method is for the
planned first release of intermediate Gaia data. As this may be
based on too short a stretch of data for a reliable five-parameter
solution, the release is targeted to give only positions and
mean $G$.\footnote{A tentative release schedule is given by ESA on
\url{http://www.cosmos.esa.int/web/gaia/release}.}
We propose the use of a prior to ensure that the one year global solution
provides a sensible formal position uncertainty for all stars.
This scenario is different compared to Sect.~\ref{sec:secondary}, since the
attitude must now be determined from the same observations as the astrometric
parameters, a so-called primary solution.\footnote{It could also be considered
to use the attitude from a potential Tycho--Gaia Astrometric Solution
\citep{2015A&A...574A.115M}. }
We simulate this through a global solution assuming one year of Gaia
observations with 20\% of dead time, and using priors of varying size.

Table~\ref{tab:results1yr} summarizes our results.
Contrary to Sect.~\ref{sec:secondary} and Table~\ref{tab:results} we now find that the prior
$\sigma_{\varpi,F90}$ constrains the solution too much. It appears that
the prior uncertainty needs to be increased to account for the
larger attitude errors caused by the unknown parallax and proper motion contributions.
Empirically we find that a ten fold increase of the prior uncertainty provides the necessary
relaxation of the constraint and allows the solution to fulfill the criteria for a sensible
astrometric result.

\section{Conclusions}
In this paper we discuss the astrometric solutions for stars with
an insufficient number of Gaia observations. This will be the case for the majority of
stars in the first data release of Gaia data, but is also an important issue
during later stages of the mission, e.g. for transient objects that are only
observed in their bright phases, and stars close to the detection limit.
In all these cases one can still obtain very valuable position estimates,
either by solving only for the position parameters or through the use of
priors for the remaining parameters. In fact, solving only for the position parameters is
equivalent to assuming that the parallax and proper motion are exactly zero,
in other words to the use of prior values equal to zero with infinite weights.
Using a more carefully selected prior improves the quality of the astrometric
solution for these stars. Very specifically, it provides
an elegant way to ensure that the position estimates obtain formal uncertainties
that correctly characterize the actual errors.

Prior information is incorporated in the astrometric solution using
Bayes' rule. For practical reasons the prior probability distributions are
taken to be Gaussian. Moreover, they are always centred on zero, since any
other choice would necessarily involve additional assumptions and thus
be even more arbitrary. For objects with negligible parallax,
such as quasars, it is a conservative choice.

We analyse the influence of different priors on the astrometric solutions,
based on numerical experiments with realistic distributions of stellar parameters
from the Gaia Universe Model Snapshot (GUMS). To optimize the prior we
require that 90\% of the actual position errors are included in the 90\% confidence
region calculated from the (Gaussian) posterior probability density, i.e.\ from the
formal covariance matrix.
Using the resulting prior ($\varpi_p=0 \pm \sigma_{\varpi,F90}$) we find that
non-singular five-parameter astrometric solutions can be obtained, with
reasonable estimates of the position uncertainties, for any star that is
observed in at least one field-of-view transit. Using this prior slightly
reduces the actual position errors, compared with a two-parameter solution.
The solution is robust to using a larger prior uncertainty than
$\sigma_{\varpi,F90}$, and in some cases (depending on the attitude estimation)
a ten fold increase is motivated (Sect.~\ref{sec:primary}).

The choice of a 90\% confidence level for the position errors is arbitrary and
it would be possible to optimize the prior for any other desired percentage.
The 10\% stars falling outside the confidence ellipse cannot easily be
identified from the data and could be considered outliers. In statistical uses
of the data, 90\% provides a good compromise between keeping a reasonably small
fraction of outliers and maintaining a good characterization of the positional
uncertainties for most stars. A higher confidence level would decrease the
fraction of outliers, but at the expense of a rapidly growing confidence region
due to the non-Gaussian nature of the actual position errors. 

Like any solution using a prior, the resulting astrometric parameters are
in general biased. Using a reference epoch centred on the observations,
the position bias is of the order of the neglected parallax, or at most a few
mas in typical cases. As discussed below this is acceptable. In order to
obtain realistic uncertainties of the positions, it is necessary to
introduce the parallax and proper motion as formal parameters in the
solution. This means that posterior estimates are also provided for these.
However, the resulting parallaxes and proper motions are in general so strongly
biased by the prior that they become physically meaningless, and they should
therefore not be used.

For the first release of Gaia data, consisting mainly of position information
and $G$ magnitudes, small biases in the resulting position estimates are fully
acceptable and unavoidable. For future releases however, where a full solution
can be determined for most of the stars, it is important to determine attitude
and geometric calibration parameters as part of the primary solution, by using
only the stars with a sufficient amount of observations. For all of these it is
mandatory that no prior is used. Any star with insufficient observations, which
requires the use of a prior for the solution, must be part of a separate
(secondary) update, in which the attitude and calibration are not modified.
For the secondary stars with insufficient data, the $\varpi_p=0 \pm \sigma_{\varpi,F90}$
prior will however allow us to obtain sensible position estimates with
well-characterized formal uncertainties.


\begin{acknowledgements}
We thank Uwe Lammers, who contributed to the idea and progress of
this study, and its funding under ESA Contract No.~4000108677/13/NL/CB.  We
are grateful to Ulrich Bastian, Alex Bombrun,
Anthony Brown, Sergei Klioner, Uwe
Lammers, Paul McMillan, Mercedes Ramos-Lerate, and the referee, Floor van
Leeuwen, for their helpful feedback to our manuscript. This work uses the
AGIS/AGISLab and GaiaTools software
packages, and we extend our special thanks to their
maintainers and developers. We gratefully acknowledge financial support from
the Swedish National Space Board and the Royal Physiographic Society in Lund.
\end{acknowledgements}

\bibliographystyle{aa} 
\bibliography{bayesianStudy} 

\appendix

\section{Analytical illustration\label{sec:appendix}}

To analytically study how the prior affects the posterior position
uncertainty, we consider a simplified case where the solution includes only two
astrometric parameters: one component of position (e.g.\ $\delta$) and the
parallax ($\varpi$). The normal matrix incorporating the prior $\vec{N}_{p} = \textrm{diag}(0,
\sigma_{\varpi,p}^{-2})$, similar to Eq.~(\ref{eq:prior}), then takes the form
\begin{equation}
  \vec{N}_0+\vec{N}_p=\frac{1}{1-\rho^2}
  \left(
 \begin{array}{cc}
  \sigma_{\delta}^{-2} & \dfrac{-\rho}{\sigma_\delta\sigma_\varpi} \\[12pt]
  \dfrac{-\rho}{\sigma_\delta\sigma_\varpi} & \sigma_\varpi^{-2}+\left(1-\rho^2\right)\sigma_{\varpi,p}^{-2}
 \end{array}
 \right)\,,
\end{equation}
where $\sigma_\delta$ and $\sigma_\varpi$ are the standard errors of the
position and parallax, respectively, and $\rho$ is the correlation coefficient;
all these quantities are based on the data only.  Calculating the posterior
covariance matrix using Eq.~(\ref{eq:posteriorC}) we find the position
uncertainty
\begin{equation}
 \label{eq:appendix}
  \sigma_{\delta,\,\textrm{posterior}}=\sigma_{\delta}\sqrt{1-\frac{\rho^2}{1+\left(\sigma_{\varpi,p}/\sigma_\varpi\right)^2}}\,.
\end{equation}
This formula agrees with our numerical experiments. In particular
it reproduces the behaviour of the position uncertainty shown in
Figs.~\ref{fig:behaviourSummary} and \ref{fig:othertransits}, featuring a
monotonically increasing $\sigma_{\delta,\,\textrm{posterior}}$ between two
asymptotic values, $\sigma_\delta\sqrt{1-\rho^2}$ and $\sigma_\delta$, as the
prior goes from very tight to very loose.
Equation~(\ref{eq:appendix}) implies that the improvement in the positional
uncertainty gained by using the parallax prior depends only on the correlation
coefficient $\rho$ and the ratio of the parallax prior to the formal
uncertainty without prior, $\sigma_{\varpi,p}/\sigma_\varpi$. For uncorrelated
data, no improvement is possible, as
$\sigma_{\delta,\,\textrm{posterior}}=\sigma_\delta$ for $\rho=0$. 

 Similar
arguments hold for the general five-parameter solution, except
that they cannot be demonstrated so easily.

\end{document}